\documentclass[unnumsec,webpdf,contemporary,large,unicode=true,psdextra]{oup-authoring-template}%
\usepackage{xcolor}
\colorlet{blue}{black}
\colorlet{red}{black}

\onecolumn 

\graphicspath{{Fig/}}

\usepackage{amsmath}
\usepackage{graphicx}
\usepackage{enumerate}
\usepackage{natbib}
\usepackage{url} 
\usepackage{epsfig}  
\usepackage{multirow} 
\usepackage{multicol} 
\usepackage{float}
\usepackage{booktabs}
\usepackage[compact]{titlesec}
\usepackage{textcomp}
\usepackage{soul}
\usepackage{tikz}
\usetikzlibrary{positioning, fit, arrows.meta}
\usetikzlibrary{bayesnet}   
\usetikzlibrary{shapes,arrows,positioning}
\usepackage{amsmath, amsthm, amssymb}
\usepackage[T1]{fontenc}

\usepackage{enumitem}
\usepackage{endfloat}
\usepackage{makecell}

\hypersetup{
  breaklinks=true,   
  colorlinks=true,
  linkcolor=black,
  citecolor=black,
  urlcolor=black,
  filecolor=black
}

\usepackage{url}  

\usepackage[normalem]{ulem}
\newcommand{\blue}[1]{\textcolor{blue}{#1}}

\theoremstyle{thmstyleone}%
\newtheorem{theorem}{Theorem}
\newtheorem{lemma}[theorem]{Lemma}
\theoremstyle{thmstyletwo}%
\theoremstyle{thmstylethree}%

\begin{document}

\journaltitle{Journal Title Here}
\DOI{DOI HERE}
\copyrightyear{2022}
\pubyear{2019}
\access{Advance Access Publication Date: Day Month Year}
\appnotes{Paper}

\firstpage{1}


\title[BRACE]{Flexible aggregation of compositional predictors with shared effects  for microbiome association analysis}

\author[1,$\ast$]{Satabdi Saha \ORCID{0000-0002-0117-9431}}
\author[2]{Liangliang Zhang}
\author[3]{\textcolor{blue}{Michele Guindani}}
\author[1]{Kim-Anh Do}
\author[4]{Christine B.\ Peterson}

\authormark{Saha et al.}
\address[1]{\orgdiv{Department of Biostatistics}, \orgname{The University of Texas MD Anderson Cancer Center}, \orgaddress{\street{7007 Bertner Avenue}, \state{Houston, TX}  \postcode{77030}, \country{USA}}}
\address[2]{\orgdiv{Department of Population and Quantitative Health Sciences}, \orgname{Case Western Reserve University}, \orgaddress{\street{ 2109 Adelbert Rd,}  \state{Cleveland, OH}  \postcode{44106}, \country{USA}}}
\address[3]{\orgdiv{Department of Biostatistics},\orgname{ UCLA Fielding School of Public Health},\orgaddress{ Los Angeles},  \state{CA} \postcode{90095}, \country{USA}}
\address[4]{\orgdiv{Department of Statistics},\orgname{ Rice University},\orgaddress{ 6100 Main St, Houston}, \state{TX}  \postcode{77030}, \country{USA}}


\corresp[$\ast$]{\textit{Address for correspondence}: Satabdi Saha, Department of Biostatistics, The University of Texas MD Anderson Cancer Center, 7007 Bertner Avenue, Houston, TX 77030,  USA. Email: \href{satabdisaha1288@gmail.com}{satabdisaha1288@gmail.com}}

\received{Date}{0}{Year}
\revised{Date}{0}{Year}
\accepted{Date}{0}{Year}



\abstract{Ongoing advancements in microbiome profiling have provided unprecedented insights into the molecular dynamics of microbial communities, sparking a surge of interest in uncovering the microbiome’s critical role in human health. Identifying microbial features linked to clinical outcomes, however, remains challenging due to the high-dimensional, sparse, and compositional nature of microbiome data. Additionally, many microbial taxa, although classified as distinct, may share functional roles, complicating traditional variable selection methods. To overcome these obstacles, we introduce Bayesian Regression with Agglomerated Compositional Effects (BRACE), a novel approach using a \textcolor{blue}{spike-and-cluster prior combining Bernoulli activity indicators, \textcolor{blue}{an Ewens exchangeable partition prior on the finite active set}, and a projection-based constrained Gaussian prior on cluster effects} to perform data-adaptive clustering and variable selection. The methodological innovation of our work lies in how we \textcolor{blue}{combine \textcolor{blue}{the Ewens partition prior} with a projection-based constrained Gaussian on the cluster atoms to enforce the sum-to-zero} constraint. BRACE groups microbial taxa with similar effects on the outcome, yielding more interpretable models while enabling effective dimension reduction. Through comprehensive simulations and a real-world application examining the influence of oral microbiome composition on insulin resistance, we demonstrate BRACE’s superior performance over existing methods, particularly in identifying key features with shared effects on outcomes.}


\maketitle

\section{Introduction}
\label{sec:Intro}
The microbiome, an incredibly diverse community of organisms, plays a pivotal role in human health and disease \citep{Pflug2012}. Recent technological advances have enabled direct, high-throughput profiling of microbial communities in human samples. Most microbiome studies rely on sequencing of the 16S ribosomal RNA (rRNA) gene, which serves as a taxonomic barcode that can be mapped to bacterial taxa, often at genus level and, when resolution permits, species level. Microbiome composition is highly heterogeneous across individuals, shaped by factors such as environment, diet, medication use, and host biology. A growing body of work has linked taxonomic and functional shifts in the microbiome to a wide spectrum of health conditions, underscoring its potential as both a biomarker and a modulator of disease processes.

\textcolor{blue}{However, the analysis of microbiome profiling data poses several challenges \citep{Peterson2024}. Because sample collection and sequencing provide only a limited snapshot of the community, observed read counts are not directly linked to the total microbial biomass. Instead, the total reads per sample mostly reflect sequencing depth and other technical factors, making the data inherently compositional \citep{Gloor2017}. Consequently, meaningful comparisons across samples rely on the relative distribution of reads across taxa rather than on raw totals. Moreover, normalizing read counts by library size to obtain relative abundances forces each sample’s taxa proportions to sum to one, which induces dependence among taxa and motivates regression approaches based on relative (typically log-ratio) information for scale-invariant interpretation.} In addition, the high dimensionality of microbiome data necessitates sparse modeling techniques to identify relevant features for the prediction of clinical outcomes \citep{lin2014variable}. Finally, microbiome data sets contain numerous rare features, for instance, those observed in less than 5\% or 10\% of subjects. Typically, these rare features are filtered out prior to downstream analysis \citep{Callahan2016}.
An alternative approach to filtering is to group features at higher taxonomic levels prior to predictive modeling. 
Taxonomic trees follow the traditional hierarchical classification \textsl{Kingdom}, \textsl{Phylum}, \textsl{Class}, \textsl{Order}, \textsl{Family}, \textsl{Genus}, \textsl{Species}. Phylogenetic trees, which can be obtained via bioinformatic pipelines, instead group features on the basis of their potential evolutionary relationships, and are often integrated into statistical analysis as an external source of information on feature similarity.

In this work, we study the role of the microbiome in shaping host phenotypes. The challenge of linking the microbiome to human health outcomes can be framed as a regression problem with compositional predictors.
In early work on compositional data analysis, \cite{aitchison1984log} proposed the linear log-contrast model for regression modeling 
with compositional covariates. Building on that idea, \cite{lin2014variable} proposed applying an $l_1$ penalty to the coefficient vector of the linear log contrast model for sparse estimation of coefficients and improved  prediction accuracy in the context of high-dimensional data. \cite{Shi2016} extended this 
work by selecting subcompositions of taxa at fixed taxonomic levels. \cite{zhang2021bayesian} proposed a Bayesian model where the compositionality constraint was 
incorporated in the prior for the
coefficient vector through a conditioning matrix with a controllable shrinkage parameter.
To encourage  joint selection of phylogenetically related features, the authors incorporated information from a known phylogenetic tree through an Ising prior. \cite{zhang2024bayesian} proposed a Bayesian compositional generalized linear model 
that incorporates the phylogenetic relatedness among taxa through a structured regularized horseshoe prior. To deal with
rare features, \cite{bien2021tree} proposed grouping finer-resolution taxa into higher
levels of taxonomic resolution by aggregating features over branches of a known 
tree and using the aggregated features in the outcome prediction.

However, these existing modeling approaches have critical limitations. The penalty-based methods proposed by \cite{lin2014variable} and \cite{Shi2016} provide point estimates of regression coefficients  that do not fully capture the uncertainty. In addition, these methods are optimized for prediction, rather than feature selection, and tend to have relatively high false positive rates \citep{zhang2021bayesian}. The Bayesian approach proposed by \cite{zhang2021bayesian} requires external information about a phylogenetic tree for estimation; this information reflects global genomic similarity between species, which may be an imperfect or noisy reflection of their functional similarity in driving an outcome of interest. Although \cite{bien2021tree} account for the presence of rare features, they similarly rely on a fixed externally defined tree to achieve feature aggregation.

In this article, we introduce Bayesian Regression with Agglomerated Compositional Effects (BRACE), a framework tailored for analyzing microbiome data that adeptly navigates inherent challenges such as its compositional structure, high dimensionality, and rare features.
\textcolor{blue}{To address the fixed-sum constraint in compositional regression, we propose a spike-and-cluster prior that combines Bernoulli activity indicators, an Ewens partition prior on the finite active set, and a projection-based constrained Gaussian prior on the cluster effects  (Figure \ref{fig:plate-diagram}). 
The Ewens partition prior is the finite exchangeable partition law associated
with the Ewens sampling formula, originally introduced as a probability model
for the partition structure of a finite sample into allelic types
\citep{Ewens1972}, and it is commonly interpreted more broadly as an
exchangeable model for species-abundance or type-abundance configurations \citep[][]{Ewens1972, Antoniak1974, Kingman1978}. The corresponding partition law  can also be obtained as the finite exchangeable partition distribution induced by a Chinese restaurant process \citep{Pitman1996, Pitman2006}. In our model, it allows active coefficients to be grouped into data-adaptive clusters sharing a common effect. The projection-based Gaussian prior then enforces the required zero-sum constraint on the induced regression coefficients, enabling posterior sampling directly within the compositionally valid parameter space via Gibbs sampling.}
Furthermore, our proposed method enables simultaneous variable selection and natural clustering of compositional microbiome profiles, effectively tackling the challenge posed by rare features.
Through clustering the coefficients of sparsely observed features, we achieve substantial dimension reduction, resulting in denser features representing groups of organisms with shared effects, ultimately enhancing the prediction of 
clinical outcomes. Our innovative approach  clusters regression coefficients in a data-adaptive manner, strategically collapsing rare features to generate denser groupings. To the best of our knowledge, BRACE stands as the first Bayesian method for high-dimensional compositional regression with flexible microbiome feature aggregation and selection. 

To motivate the methodological developments proposed, we analyze the Oral Infections, Glucose Intolerance, and Insulin Resistance Study (ORIGINS)  \citep{demmer2015periodontal}, which seeks to characterize the association between the bacterial population of subgingival plaque and \textcolor{blue}{fasting levels of insulin in blood, a continuous marker used as a diagnostic measure of insulin resistance and}
prediabetes. Oral microorganisms play a key role in shaping the risk of periodontal diseases, including periodontitis \citep{pihlstrom2005periodontal}.
It is postulated that chronic inflammation driven by the periodontal microbiota may contribute to impaired glucose regulation and heightened risk of insulin resistance, \textcolor{blue}{which is characterized by excess insulin production and correspondingly higher levels of insulin in blood}, potentially laying the groundwork for type 2 diabetes \citep{gurav2012periodontitis}. Through our case study on the ORIGINS data, we aim to identify taxa associated with insulin resistance, shedding light on the influence of the periodontal microbiome on prediabetes.

\begin{figure}[ht]
    \centering
    \includegraphics[width=1\textwidth]{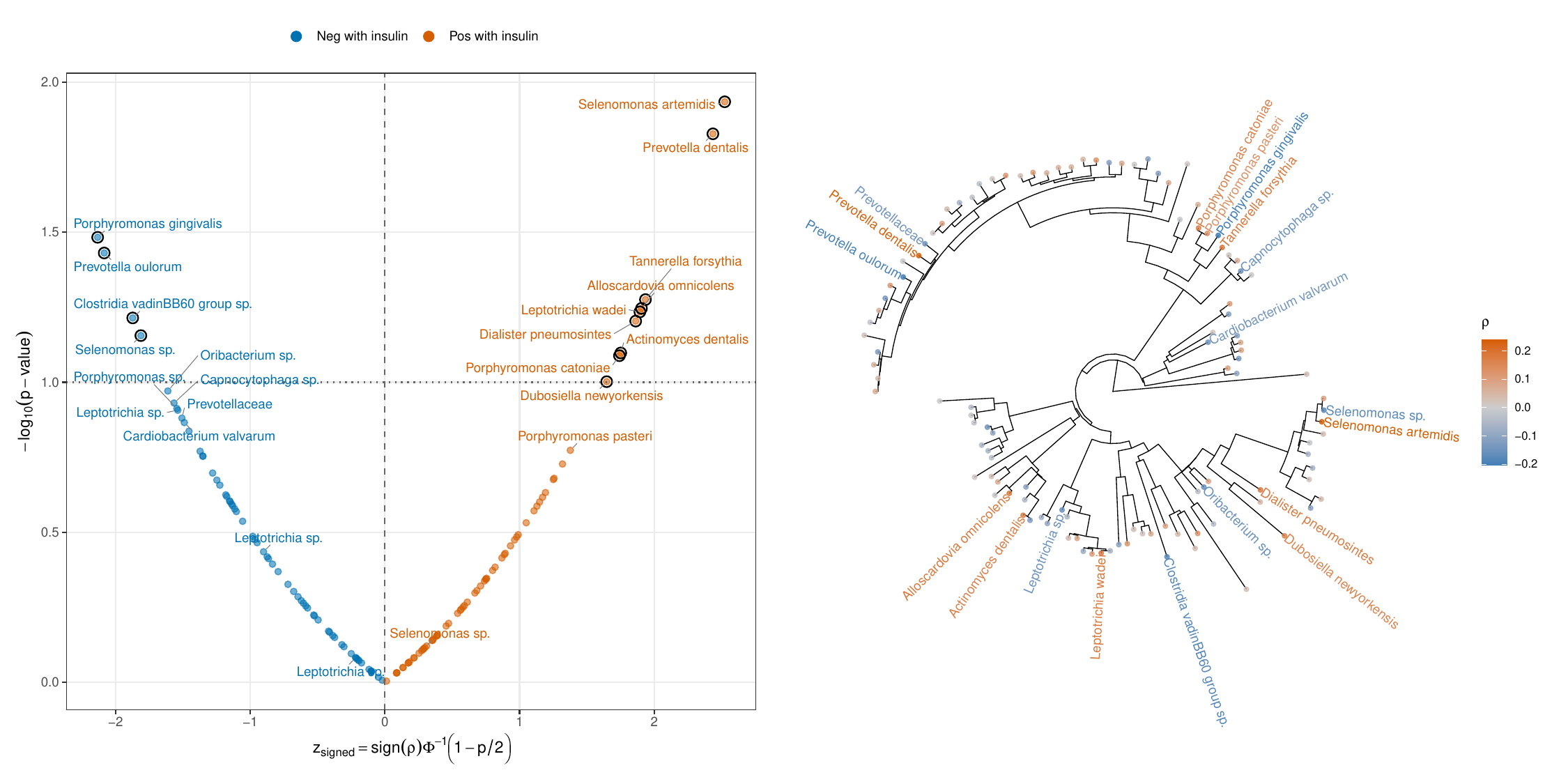}
    \caption{Species-level associations with fasting insulin are not phylogenetically structured. (A) Volcano plot of Spearman correlations between species-level relative abundances and fasting insulin levels. Each point represents a single species; the x-axis displays a signed $z$-score ($z_{\text{signed}}$ = sign($\rho$) $\cdot$ $\Phi^{-1}$(1 - $p/2$)) and the y-axis shows $-\log_{10}$(p-value). Orange and blue points denote positive and negative correlations, respectively. The horizontal dotted line marks p = 0.10. Black rings highlight the top 10 positively and top 10 negatively correlated species by $\rho$.
    (B) Phylogenetic tree labeled for the same 20 species, with tip colors indicating direction of association. Closely related taxa frequently display opposing associations, for example, \emph{Selenomonas artemidis} (positive) and \emph{Selenomonas} sp.\ (negative) are phylogenetic neighbors yet fall on opposite sides of the association spectrum.}
    \label{fig:motivation_fig}
\end{figure}

\textcolor{blue}{One of our primary goals is data-adaptive feature aggregation, motivated by the concern that aggregation based solely on an external phylogenetic tree may not align with taxa effects on the outcome of interest.
To examine this in the ORIGINS dataset, we first computed Spearman rank correlations between insulin levels and the relative abundance of each detected species.
We then selected the 10 most positively and 10 most negatively correlated species and visualized them on the phylogenetic tree.
This structure did not reveal coherent clustering by association direction: closely related taxa including  pairs such as \emph{Selenomonas artemidis} and \emph{Selenomonas} sp., which belong to the same genus, exhibited correlations with insulin with differing signs (Figure \ref{fig:motivation_fig}).
To quantify this mismatch more formally, we evaluated phylogenetic signal in the vector of $\rho$ values across the full tree using Pagel's $\lambda$ and Blomberg's $K$, which measure the tendency of closely related species to exhibit similar trait values \cite{pagel}.
Both statistics indicated negligible phylogenetic signal ($\lambda = 0.0001$, $p=1$; $K=0.049$, $p=0.82$), confirming that the strength  and direction of species–insulin associations are not congruent with phylogenetic structure. These findings suggest that the metabolic relevance of individual oral taxa is shaped by species-specific functional traits rather than shared evolutionary history, and that phylogenetic proximity alone should not be used to guide association analysis in this community.}

The remainder of the article is organized as follows: In Section \ref{sec:methods}, we provide a description of the proposed model and estimation procedure. In Section \ref{sec:simulation}, we benchmark the performance of BRACE against alternative approaches through simulation studies. In Section \ref{sec:case_study}, we demonstrate the applicability of BRACE through a study that aims to capture the relationship between subgingival microbial
community composition and \textcolor{blue}{fasting insulin levels}. Finally, Section \ref{sec:conclusion} includes a discussion and concluding remarks. 

\begin{figure}[ht]
\centering
\begin{tikzpicture}[
  >=Latex,
  node distance=1.45cm and 2.10cm,
  line width=0.45pt,
  shorten >=1pt, shorten <=1pt,
  every node/.style={font=\small},
  obs/.style   ={circle, draw, fill=gray!25, minimum size=18pt, inner sep=1pt},
  latent/.style={circle, draw, fill=white,   minimum size=18pt, inner sep=1pt},
  const/.style ={inner sep=0pt},
  fac/.style   ={rectangle, fill=black, minimum size=6pt, inner sep=0pt},
  flab/.style  ={font=\scriptsize, inner sep=0pt},
  plate/.style ={draw, rounded corners, inner sep=7pt}
]

\node[obs] (y) {$y_i$};
\node[fac, above=3mm of y] (yfac) {};
\node[flab, left=2mm of yfac, yshift = 1mm] {$\mathcal N$};
\node[obs, left=2.2cm of y] (x) {$\mathbf X_i$};

\node[latent, above right=1.0cm and 1.7cm of yfac] (sig) {$\sigma^2$};
\node[fac, above=3mm of sig] (sigfac) {};
\node[flab, left=2mm of sigfac] {IG};
\node[const, above=4mm of sigfac] (sigpar) {$a_\sigma,b_\sigma$};

\node[latent, below=1.95cm of y] (beta) {$\beta_j$};

\node[fac, above=3mm of beta, xshift= 6mm] (det) {};
\node[flab, right=3mm of det] {Det};

\node[latent, left=2.45cm of beta] (s) {$s_j$};
\node[fac, above= 2mm of s] (sfac) {};
\node[flab, left = 2mm of sfac, yshift= 1mm ] {Bern};

\node[latent, left=1.95cm of s] (psi) {$\psi_0$};
\node[fac, above=3mm of psi] (psifac) {};
\node[flab, left = 2mm of psifac] {Beta};
\node[const, above=4mm of psifac] (a0) {$\alpha_0$};

\node[latent, below=1.55cm of beta] (z) {$z_j$};
\node[fac, left=5mm of z] (zfac) {};
\node[flab, above = 1mm of zfac, yshift= -6mm] {Ewens};

\node[latent, left=1.95cm of z] (alp) {$\alpha$};
\node[fac, below=3mm of alp] (alfac) {};
\node[flab, left=2mm of alfac] {Ga};
\node[const, below=4mm of alfac] (alppar) {$a_\alpha,b_\alpha$};

\node[latent, right=2.75cm of beta] (theta) {$\theta_k$};
\node[fac, above=3mm of theta] (thfac) {};
\node[flab, left=2mm of thfac, xshift =  2mm] {$\mathcal N_{\mathcal C}$};
\node[const, above=4mm of thfac, align=center] (Pw)
  {$\mathbf P_w$\\[-1pt]\scriptsize$\boldsymbol w^\top\boldsymbol\theta=0$};

\node[latent, right=1.95cm of theta] (gam) {$\gamma^2$};
\node[fac, above=3mm of gam] (gamfac) {};
\node[flab, right=2mm of gamfac] {IG};
\node[const, above=4mm of gamfac] (gampar) {$a_\gamma,b_\gamma$};

\draw[->] (x) -- (yfac);
\draw[->] (beta) -- (y);
\draw[->] (yfac) -- (y);

\draw[->] (sigpar) -- (sigfac);
\draw[->] (sigfac) -- (sig);
\draw[->] (sig) to[out=225,in=40] (yfac);

\draw[->] (a0) -- (psifac);
\draw[->] (psifac) -- (psi);
\draw[->] (psi) -- (sfac);
\draw[->] (sfac) -- (s);

\draw[->] (alppar) -- (alfac);
\draw[->] (alfac) -- (alp);
\draw[->] (alp) -- (zfac);
\draw[->] (zfac) -- (z);

\draw[->] (Pw) -- (thfac);
\draw[->] (gampar) -- (gamfac);
\draw[->] (gamfac) -- (gam);
\draw[->] (gam) -- (thfac);
\draw[->] (thfac) -- (theta);

\draw[->] (s)     to[bend left=18]  (det);
\draw[->] (z)     to[bend right=18] (det);
\draw[->] (theta) to[bend left=12]  (det);
\draw[->] (det) -- (beta);

\draw[->] (s) to[out=-55,in=160] (z);

\node[plate, fit=(x)(y)(yfac), label=below right:{\scriptsize $i=1,\ldots,n$}] {};
\node[plate, fit=(s)(sfac)(beta)(det)(z)(zfac), label=below right:{\scriptsize $j=1,\ldots,p$}] {};
\node[plate, fit=(theta)(thfac), label=below right:{\scriptsize $k=1,\ldots,K$}] {};

\end{tikzpicture}
\caption{Plate diagram for BRACE, Bayesian Regression with Agglomerated Compositional Effects. The model combines three components: (i)~Bernoulli spike indicators $s_j$ for variable selection, 
    (ii)~an Ewens exchangeable partition prior on active indices for data-adaptive feature aggregation, 
    and (iii)~a projection-based constrained Gaussian prior on the cluster atoms $\boldsymbol{\theta}$ 
    to enforce the compositional constraint $\mathbf{1}^\top\boldsymbol{\beta} = 0$. Shaded circles are observed; open circles are latent; filled squares are factor nodes. }
\label{fig:plate-diagram}
\end{figure}

\section{Methods} \label{sec:methods}
\subsection{Background}
\noindent

BRACE builds on the  compositional regression framework
for the prediction of continuous outcomes from microbiome profiles. Let $\boldsymbol{y} = \{y_1, \ldots y_n\}$ denote the vector of continuous responses across $n$ samples, \textcolor{blue}{$\boldsymbol{y}\in\mathbb{R}^n$}, and $\mathbf{U}$ denote the $n \times p$ observed microbial abundance matrix for $p$ features. 
Notably, the sequencing methods used for generating the microbial abundances result in compositional data. \textcolor{blue}{Because sequencing depth varies across samples, microbial read counts are constrained by a sample-specific total (library size) and are therefore identifiable only on a relative scale.} 
Prior to downstream analysis, the observed abundance tables are generally converted to relative abundance matrices using a data transformation. \textcolor{blue}{To mitigate numerical instability arising from zero counts, small pseudocounts are commonly introduced prior to log-ratio transformation or normalization. Two common approaches are generally used: (i) zero-replacement strategies, in which only zero entries are replaced with a small value (e.g., half the minimum observed nonzero abundance) \cite{lin2014variable,zhang2021bayesian,zhang2024bayesian} and (ii) additive pseudocount strategies, in which a constant is added to all entries in the count matrix \cite{bien2021tree,shi2022high}.} To obtain the relative abundances, we rely on total sum scaling, where each element of the count matrix $u_{ij}$ is divided by its sample sum $\sum_{j=1}^p{u_{ij}}$.  The resulting relative abundance matrix $\tilde{\mathbf{U}}$ 
 satisfies the compositional constraint $\sum_{j=1}^p \tilde{u}_{ij} = 1$. Each row of $\tilde{\mathbf{U}}$  is constrained to the simplex $\mathcal{S}^{p}$, rather than the unrestricted real space $\mathbb{R}^p$. 
$\tilde{\mathbf{U}}$ is then log transformed to obtain the log relative abundance matrix $\mathbf{X}$, where $\mathbf{X} = \log(\tilde{\mathbf{U}})$, \textcolor{blue}{$\mathbf{X}\in\mathbb{R}^{n\times p}$}. Importantly, the $p$ features in the matrix $\mathbf{X}$ are still dependent due to the original compositionality constraint. 


To deal with compositionally constrained covariates in the regression framework, \cite{aitchison1984log} proposed the linear log-contrast model: 
\begin{align}
    \label{eqn:ALRmodel}
    \boldsymbol{y} = \mathbf{C}_{\char`\\ p}\boldsymbol{\beta}_{\char`\\ p} + \boldsymbol{\varepsilon},
\end{align} 
where $\mathbf{C}_{\char`\\ p}= \log(u_{ij}/u_{ip})$ is an $n \times (p-1)$ matrix of the additive log-ratio (ALR) transformed predictor variables, with the transformation done using the $p^{\text{th}}$ predictor as the reference component, $\boldsymbol{\beta}_{\char`\\ p} = \{\beta_1,\ldots,\beta_{p-1}\}$ is the vector of $p-1$ regression coefficients, and the entries in the noise vector are distributed independently as $\varepsilon_i \sim \mathcal{N}(0, \sigma^2)$, for $i = 1, \ldots, n$.  {\color{blue}Letting $X_{ij}=\log(u_{ij})$ gives ${\mathbf{C}_{\char`\\ p}}_{ij}=X_{ij}-X_{ip}$, so $\mathbf{C}_{\char`\\ p}\boldsymbol{\beta}_p=\mathbf{X}\boldsymbol{\beta}$ with $\beta_j={\beta_{\char`\\ p}}_{j}$ for $j=1,\ldots,p-1$ and $\beta_p=-\sum_{j=1}^{p-1}{\beta_{\char`\\ p}}_{j}$, which implies $\boldsymbol{1}^\top\boldsymbol{\beta}=0$. This motivates a reference-free but equivalent parameterization \citep{lin2014variable}:
\begin{align}
\label{constrainedLR}
    \boldsymbol{y} = \mathbf{X}\boldsymbol{\beta}+ \boldsymbol{\varepsilon} \hspace{0.1in} \quad \text{subject to the constraint} \hspace{0.1in} \boldsymbol{1^{\top}\beta} = 0.
\end{align}
The intercept term is omitted by centering the response and predictor variables. The zero-sum constraint $\mathbf{1}^\top\boldsymbol{\beta}=0$ ensures that the linear predictor $\mathbf{X}_i^\top\boldsymbol{\beta}$ depends only on relative abundances and is invariant to sample-specific shifts of the log-covariates: for any constant $a_i$, $(\mathbf{X}_i+a_i\mathbf{1})^\top\boldsymbol{\beta}=\mathbf{X}_i^\top\boldsymbol{\beta}$. Consequently, coefficients are interpretable only through contrasts: differences $\beta_j-\beta_k$ quantify the association with the log-contrast $X_{ij}-X_{ik}$, so increasing $X_{ij}-X_{ik}$ by one unit shifts the mean outcome by $(\beta_j-\beta_k)$. This reference-free formulation is therefore equivalent to the ALR model while avoiding dependence on a chosen denominator component. Additional characterization showing how the sum-to-zero constraint naturally accommodates the commonly used additive log-ratio (ALR) and centered log-ratio (CLR) transformations is provided in Supplementary Section S3.}

Our goal is to make inference on the regression coeﬃcients $\boldsymbol{\beta}$ based on the observed data $(\boldsymbol{y}, \mathbf{X})$ where it is assumed that any prior on $\boldsymbol{\beta}$ satisfies $Pr( \mathcal{S}) = 1$, where $\mathcal{S} = \{\boldsymbol{\beta}: \boldsymbol{1^{\top}\beta} = 0\}$. It is important to note that when the model coefficients should satisfy the constraint $\boldsymbol{1^{\top}\beta} = 0$, using unrestricted priors on $\boldsymbol{\beta}$ (such as a standard multivariate normal or continuous shrinkage priors) can lead to a loss of finite sample efficiency for the parameter estimates by disregarding this structural constraint. In addition, the posterior distribution of $\boldsymbol{\beta}$ might assign probability mass to regions of the parameter space that are not feasible under the true model, which will ultimately lead to invalid parameter estimates and violation of the constraint space. An illustrative example with two predictors is provided in  Supplementary Section S1.

Another important consideration in modeling microbiome data is the presence of highly sparse features. To improve signal, a common approach is to group finer-resolution taxa into higher levels of taxonomic resolution by summing over all the features that belong to the corresponding classification in a known taxonomic tree. More generally, suppose we obtain a new aggregated feature for the $i$th subject $x_{i,a} = x_{i,1} + x_{i,2} + \cdots + x_{i,m}$, where $m$ denotes the number of leaf nodes descending from the ancestor node $a$. As noted in \cite{yan2021rare}, in the linear model setting, $x_{i,a}\beta = (x_{i,1} + x_{i,2} + \cdots x_{i,m})\beta = x_{i,1}\beta  + x_{i,2}\beta + \cdots x_{i,p}\beta$. Effectively, this means that learning a model where some features have exactly equal coefficients $\beta$ corresponds to aggregation of the original rare features into a smaller set of more common features. A key limitation of \cite{bien2021tree} is that they assume that this aggregation must occur within branches of a known tree structure; this assumption is overly rigid, as any external phylogenetic/taxanomic tree structure will not precisely reflect the target of interest, which is similarity of effects on the regression outcome. 

\textcolor{blue}{We instead propose to employ a prior distribution on $\boldsymbol{\beta}$ that effectively constrains the parameter space while allowing for feature aggregation. To achieve simultaneous feature selection and data-adaptive clustering, in the next section we describe a prior construction that combines an Ewens partition prior on the active predictors with jointly constrained Gaussian atoms on the model coefficients. This strategy allows features with similar impacts on the outcome to share common regression effects, providing a flexible mechanism for effect-level aggregation. Additionally, the spike component imposes sparsity, while the partition prior clusters the nonzero coefficients. We now provide the mathematical formulation of our proposed model.}


{\color{blue}\subsection{Model specification}

We specify a hierarchical Bayesian model, BRACE, 
that simultaneously performs (i) variable selection, (ii) \textcolor{blue}{data-adaptive}
clustering of active coefficients, and (iii) enforces the log-contrast 
identifiability constraint intrinsic to compositional regression. The 
hierarchy is built in three stages: a spike-and-slab layer selects an 
active set; an Ewens partition prior on the active set induces clusters of coefficients sharing a common effect; and a constrained 
Gaussian prior on the cluster-level effects enforces the log-contrast 
constraint. We describe each stage in turn and then state the induced 
marginal prior on the regression coefficients in 
Proposition~\ref{prop:beta-prior}.
\textcolor{blue}{More in detail, we model
\begin{equation}
\label{eq:likelihood}
\boldsymbol y \mid \mathbf X, \boldsymbol\beta,\sigma^2 \sim \mathcal{N}(\mathbf X \boldsymbol\beta,\sigma^2 \mathbf{I}_n), \qquad \sigma^2>0,
\end{equation}
subject to the log-contrast constraint $\boldsymbol{1}_p^\top \boldsymbol\beta = 0$. 
In order to perform variable selection, we introduce a binary activity indicator $s_{j}$ for each predictor $j=1,\ldots,p$. Predictors with $s_{j}=1$ are included in the active set, whereas predictors with $s_{j}=0$ are assigned to the spike, so that their corresponding coefficients are set exactly to zero. More specifically, we assume 
\begin{equation}
\label{eq:selection}
s_j \stackrel{\mathrm{iid}}{\sim} \mathrm{Bernoulli}(\psi_0) \qquad 
\psi_0 \sim \mathrm{Beta}\!\left(\frac{\alpha_0}{2},\frac{\alpha_0}{2}\right).
\end{equation}
Let
\(
A(\boldsymbol{s})=\{j:s_j=1\}
\)
denote the set of active predictors, and let 
\(m=|A(\boldsymbol{s})|\) be
the number of active predictors. Conditional on the active set \(A(\boldsymbol{s})\), we place an exchangeable partition prior on the active indices, allowing active coefficients to be grouped into clusters sharing a common effect. Specifically, let $\boldsymbol z=\{z_j\}_{j\in A(\boldsymbol{s})}$ denote the resulting
cluster labels, where $z_j\in\{1,\ldots,K\}$, and let
$w_k=|\{j\in A(\boldsymbol{s}):z_j=k\}|$
denote the number of active predictors assigned to cluster $k$. Then, we assign the
active indices an Ewens partition prior, 
\begin{equation*}
p(w_1, \ldots, w_K\mid \boldsymbol s,\alpha)
=
\frac{\alpha^K}{\alpha^{(m)}}\prod_{k=1}^K (w_k-1)!
\;\propto\;
\alpha^K\prod_{k=1}^K (w_k-1)!,
\end{equation*}
where $\alpha^{(m)}=\alpha(\alpha+1)\cdots(\alpha+m-1)$ and $\alpha>0$. This prior can be equivalently obtained as the finite-dimensional exchangeable partition probability function (EPPF) induced by a Chinese restaurant process (CRP) with concentration parameter $\alpha$ \citep{Pitman1996}. Therefore, we can use the CRP predictive rules as a convenient sequential representation of the finite partition prior.}
 \textcolor{blue}{The concentration parameter \(\alpha\) controls the prior tendency to form clusters among the active predictors, with larger values favoring a larger number of clusters. To map the cluster-level effects back to the original predictor space, we define the cluster membership matrix $\mathbf{Z}\in\{0,1\}^{p\times K}$, with elements
\(
Z_{jk}=\mathbb{I}\{j\in A(\boldsymbol{s}),\, z_j=k\}.
\) 
Thus, if predictor $j$ is active and assigned to cluster $k$, then $Z_{jk}=1$; otherwise $Z_{jk}=0$. In particular, rows corresponding to inactive predictors, $j\notin A(\boldsymbol{s})$, are identically zero.} 

\blue{Let
\(
\boldsymbol{w} = \mathbf Z^\top \boldsymbol 1_p = (w_1,\ldots,w_K)^\top
\)
denote the vector of cluster sizes, i.e.,  the number of predictors sharing each cluster effect, and let  $\boldsymbol{\theta}=(\theta_{1},\ldots,\theta_{K})^{\top}$ denote 
the cluster-specific effects. Given $(\boldsymbol{s}, \boldsymbol{z}, \boldsymbol{\theta})$, the full coefficient vector is then defined as
$
\boldsymbol\beta=\mathbf Z\boldsymbol\theta.
$
Under this representation, inactive predictors have coefficient zero, while
all active predictors assigned to the same cluster share the same coefficient.} 

\textcolor{blue}{It remains to enforce the log-contrast constraint
$\boldsymbol{1}_{p}^{\top}\boldsymbol{\beta}=0$. Since 
$\boldsymbol\beta=\mathbf Z\boldsymbol\theta$, it follows that $\mathbf 1_p^\top\boldsymbol\beta=(\mathbf Z^\top\mathbf 1_p)^\top\boldsymbol\theta
$; thus, the constraint can
be written at the cluster level as the size-weighted constraint 
\(\boldsymbol w^\top\boldsymbol\theta=0\). To impose the size-weighted constraint, define 
\begin{equation*}
\mathbf{P}_{w}
=\mathbf{I}_{K}-\frac{\boldsymbol{w}\boldsymbol{w}^{\top}}{\boldsymbol{w}^{\top}\boldsymbol{w}}.
\end{equation*}
This is the orthogonal projection onto the hyperplane 
$\mathcal{H}_{w}=\{\boldsymbol{\theta}\in\mathbb{R}^{K}:\boldsymbol{w}^{\top}\boldsymbol{\theta}=0\}$, so projecting an unconstrained vector by \(\mathbf P_w\) removes its component in the direction of \(\boldsymbol w\) and yields a vector satisfying \(\boldsymbol w^\top\boldsymbol\theta=0\).
Conditional on $(\boldsymbol s, \boldsymbol z,\gamma^2)$, if $K=0$ we set $\boldsymbol \beta=0$. If $K\ge 1$,  we first draw
$
\widetilde{\boldsymbol\theta}\mid \gamma^2
\sim
\mathcal N_K(\boldsymbol 0,\gamma^2\mathbf I_K),
$
and then define the cluster effects by projection, by setting
$
\boldsymbol\theta=\mathbf P_w\widetilde{\boldsymbol\theta}.
$}
\textcolor{blue}{Equivalently, $\boldsymbol{\theta}$ is distributed according to the
degenerate Gaussian measure obtained by projecting an isotropic Gaussian
vector onto $\mathcal H_w$, i.e.,
\begin{equation}
\label{eq:degenerate}
\boldsymbol \theta \mid \boldsymbol s, \boldsymbol z,\gamma^2 \sim \mathcal{N}_K(0,\gamma^2 \mathbf P_w), 
\end{equation}
a singular Gaussian measure supported on \(\mathcal H_w\), not a full-rank Gaussian density on \(\mathbb R^K\). We complete the prior specification with
\begin{equation}
\label{eq:hyperpriors}
\sigma^{2}\sim\mathrm{IG}(a_{\sigma},b_{\sigma}),
\qquad
\alpha\sim\mathrm{Gamma}(a_{\alpha},b_{\alpha}).
\end{equation}
The construction in \eqref{eq:selection}--\eqref{eq:degenerate} induces a structured prior on $\boldsymbol{\beta}$ with three main properties: exact sparsity through the spike component, equality of nonzero coefficients within data-adaptive clusters, and almost-sure support on the log-contrast constraint space. The following
proposition makes these properties precise.}

\textcolor{blue}{
\begin{lemma}[Induced prior on $\boldsymbol{\beta}$]
\label{prop:beta-prior}
Under the hierarchical model  \eqref{eq:likelihood}--\eqref{eq:degenerate}, the following hold.
\begin{enumerate}
\item[ (i) ] Conditional on $\left(s, z, \gamma^{2}\right)$ with $K \geq 2$, the regression coefficient vector satisfies
\begin{equation}
\label{eq:beta-prior}
\boldsymbol{\beta}\mid\boldsymbol{s},\boldsymbol{z},\gamma^{2}
\;\sim\;
\mathcal{N}_{p}\!\left(\boldsymbol{0},\,
\gamma^{2}\,\mathbf{Z}\mathbf{P}_{w}\mathbf{Z}^{\top}\right),
\end{equation}
where the Gaussian measure is singular and supported on the linear
subspace
\begin{equation*}
L(\boldsymbol{s},\boldsymbol{z})
=\Big\{\boldsymbol{\beta}\in\mathbb{R}^{p}:\;
\beta_{j}=0\ \forall j\notin A(\boldsymbol{s}),\;
\beta_{j}=\beta_{j'}\text{ whenever } j,j'\in A(\boldsymbol{s}), z_{j}=z_{j'},\;
\boldsymbol{1}_{p}^{\top}\boldsymbol{\beta}=0\Big\}.
\end{equation*}
Thus, conditional on the active set and its partition, the prior enforces exact sparsity, equality of coefficients within clusters, and the log-contrast constraint.
\item[ (ii) ] If $K=0$, then $A(\boldsymbol{s})=\emptyset$ and $\boldsymbol{\beta}=\mathbf{0}$ by construction. If $K=1$, the constraint $\boldsymbol{w}^{\top}\boldsymbol{\theta}=w_{1}\theta_{1}=0$ forces $\theta_{1}=0$, and hence $\boldsymbol{\beta}=\boldsymbol{0}$ almost surely. Therefore, nontrivial configurations require $K\ge 2$ active  clusters.
\item[ (iii) ] The joint prior on 
$(\boldsymbol{s},\boldsymbol{z},\boldsymbol{\theta},\boldsymbol{\beta})$  given $(\psi_{0},\alpha,\gamma^{2})$ factorizes as
\begin{equation}
\label{eq:joint-prior}
\begin{aligned}
p(\boldsymbol{s},\boldsymbol{z},\boldsymbol{\theta},\boldsymbol{\beta}
\mid\psi_{0},\alpha,\gamma^{2})
=\;&\left[\prod_{j=1}^{p}\psi_{0}^{s_{j}}(1-\psi_{0})^{1-s_{j}}\right]
p(\boldsymbol{z}\mid\boldsymbol{s},\alpha)\times\,\mathcal{CN}_{K}\!\left(\boldsymbol{\theta}\mid\boldsymbol{0},
\gamma^{2}\mathbf{I}_{K};\,\boldsymbol{w}^{\top},0\right)
\,\delta_{\boldsymbol{Z} \boldsymbol{}\theta}(\boldsymbol{\beta}),
\end{aligned}
\end{equation}
where 
$\mathcal{CN}_{K}(\,\cdot\mid\boldsymbol{0},\gamma^{2}\mathbf{I}_{K};\,\boldsymbol{w}^{\top},0)$ 
denotes the constrained Gaussian measure on $\mathcal{H}_{w}=\left\{\boldsymbol{\theta} \in\mathbb R^K: \boldsymbol{w}^{\top} \boldsymbol{\theta}=0\right\}$, and $\delta_{\mathbf{Z} \boldsymbol{\theta}}(\boldsymbol{\beta})$ denotes the point mass enforcing the deterministic relation $\boldsymbol{\beta}=\mathbf{Z} \boldsymbol{\theta}$.
\end{enumerate}
\end{lemma}
}
\noindent
A proof of Lemma~\ref{prop:beta-prior}, together with a full 
derivation of the induced prior and a characterization of its support,  is given in Supplementary Section~S2.

\subsection{Posterior inference} 
\label{sec:sampling}
To perform posterior inference for our proposed model, we employ a  Gibbs sampler to iteratively draw samples from the posterior full conditional distributions of the parameters. Our novel sampler addresses key technical challenges; in particular, we efficiently generate samples that satisfy the summation constraint on the model coefficients by leveraging a recent method for simulating hyperplane-truncated multivariate normal distributions \cite{cong2017fast}. Next, we describe the Gibbs sampling steps for our model.

\subsubsection{Sampling of the cluster labels \texorpdfstring{$\boldsymbol{z}$}{z}}

\textcolor{blue}{For computational convenience,  we augment the allocation variable $\boldsymbol z$ so that it encodes both variable selection and clustering. Specifically, we set
$z_j=0$ when predictor $j$ is inactive, corresponding to $s_j=0$, and set
$z_j=k\geq 1$ when predictor $j$ is active and assigned to active cluster
$k$. Thus, $z_j=0$ represents the spike at zero, while positive values of
$z_j$ encode membership in one of the nonzero coefficient clusters. Under this augmented representation, the likelihood can be rewritten as
$\boldsymbol y \mid \mathbf {X,Z} , \boldsymbol\theta,\sigma^2 \sim \mathcal{N}(\mathbf {XZ} \boldsymbol\theta,\sigma^2 I_n)$. Conditional on $\boldsymbol z$, the matrix $\mathbf Z$ is determined, and
only the $K$ nonzero cluster effects in $\boldsymbol\theta$ need to be
sampled.}

\textcolor{blue}{We update the
allocation labels one at a time. For a proposed assignment $z_j=k$, let
$\boldsymbol z^{*}$ denote the allocation vector obtained from
$\boldsymbol z$ by setting the $j$th label equal to $k$, while keeping all
other labels fixed; that is, $\boldsymbol z^{*}=(z_j^{*},\boldsymbol z_{-j})$, where
$z_j^{*}=k$ and $z_\ell^{*}=z_\ell$ for all $\ell\neq j$. The full conditional probability of this proposed
assignment is}
\begin{align}
\label{eq:sample_z_fixed}
P(z_j=k\mid \boldsymbol z_{-j},\gamma^2,\sigma^2,\alpha, \alpha_0, \mathbf{ X},\boldsymbol y)
~\propto~
P(z_j=k\mid \boldsymbol z_{-j},\alpha, \alpha_0)\;
f(\boldsymbol y\mid \gamma^2,\sigma^2,\mathbf{ X},\boldsymbol z^{*}).
\end{align}
To facilitate simultaneous feature aggregation and selection, we consider \textcolor{blue}{three possible assignments for $z_j$}: (i) the spike $z_j=0$, (ii) an existing nonzero cluster $k\in\{1,\dots,K\}$, or (iii) a new nonzero cluster.
We write $p=p_0+p_z$ with $p_0=\sum_j\mathbb I(z_j=0)$,  $p_z=\sum_j\mathbb I(z_j\ge 1)$, and let
$m_{-j,0}=\#\{l\neq j:z_l=0\}$,  $m_{-j,k}=\#\{l\neq j:z_l=k\}$.
and $
p_{-j,z}
:= \sum_{l\neq j}\mathbb I(z_l\ge 1)
= p_z - \mathbb I(z_j\ge 1)
= \sum_{k\ge 1} m_{-j,k}.
$
\textcolor{blue}{Then, integrating out $\psi_0$, we obtain the collapsed spike-Ewens, or equivalently spike-CRP, predictive probabilities as}
\begin{align}
\pi_{j0}
= \Pr(z_j=0 \mid z_{-j},\alpha_0)
&= \frac{m_{-j,0}+\alpha_0/2}{\,p-1+\alpha_0\,},\\[4pt]
\Pr(z_j=k \mid z_{-j},\alpha, \alpha_0)
&= (1- \pi_{j0})\,\frac{m_{-j,k}}{\,p_{-j,z}+\alpha\,},
\qquad k=1,\ldots,K_{-j},\\[4pt]
\Pr(z_j=\text{new} \mid z_{-j},\alpha, \alpha_0)
&= (1- \pi_{j0})\,\frac{\alpha}{\,p_{-j,z}+\alpha\,}.
\label{eq:spike_crp_fixed}
\end{align}

\subsubsection{Collapsed update for \texorpdfstring{$\boldsymbol z$}{z}: marginalizing the atoms}

\textcolor{blue}{To evaluate the single-site label updates in \eqref{eq:sample_z_fixed}, we
integrate out the $K$ nonzero cluster effects $\boldsymbol\theta$ subject  to the constraint $\boldsymbol w^\top\boldsymbol\theta=0$. This yields a collapsed marginal likelihood for each proposed allocation $\boldsymbol z^*$, so that the label update depends only on the prior predictive probability of the proposed assignment and on the fit of the model after marginalizing over the constrained cluster effects. Since the constraint removes one degree of freedom, the integration is over a $(K-1)$-dimensional hyperplane rather than over the full $K$-dimensional space, which improves computational stability and mixing. We define the cluster-collapsed design matrix
$\mathbf X_{\boldsymbol z}=\mathbf X\mathbf Z$ and set $\boldsymbol b=\mathbf X_{\boldsymbol z}^\top\boldsymbol y$, 
$
\mathbf A = \mathbf X_{\boldsymbol z}^\top \mathbf X_{\boldsymbol z}
+ \frac{\sigma^2}{\gamma^2}\,\mathbf I_K.
$} 
We  then partition $\mathbf A$ and $\boldsymbol b$ as
$\mathbf A=\begin{pmatrix}\mathbf A_{11}&\mathbf A_{12}\\ \mathbf A_{12}^\top & a_{KK}\end{pmatrix}$,
$\boldsymbol b=(\boldsymbol b^\ast,b_K)$, where $\mathbf A_{11}$ is $(K-1)\times(K-1)$ and
$\boldsymbol b^\ast$ contains the first $K-1$ components of $\boldsymbol b$. We then parameterize the hyperplane characterized by the constraint $\boldsymbol w^\top\boldsymbol\theta=0$ via the linear map $\boldsymbol\theta=\mathbf T\boldsymbol\theta^*$, where
$\mathbf T=\big(\mathbf I_{K-1},\,-w_K^{-1}\boldsymbol w^{*\top}\big)^\top$
and $\boldsymbol\theta^*\in\mathbb R^{K-1}$, so that $\theta_K=-(1/w_K)\sum_{k=1}^{K-1}w_k\theta_k$. The induced metric on the hyperplane is
$\mathbf B=\mathbf T^\top\mathbf T=\mathbf I_{K-1}+(\boldsymbol w^\ast \boldsymbol w^{\ast\top})/{w_K^2}$,
so that the Hausdorff measure on the hyperplane transforms as $d\mathcal H^{K-1}(\boldsymbol\theta)=\sqrt{\det(\mathbf B)}\,d\boldsymbol\theta^\ast$.
Using the Weinstein–Aronszajn identity, we can write
\[
\det(\mathbf B)
= \det\!\Big(\mathbf I_{K-1}+\frac{\boldsymbol w^\ast \boldsymbol w^{\ast\top}}{w_K^2}\Big)
= 1+\frac{\|\boldsymbol w^\ast\|^2}{w_K^2}
= \frac{\sum_{k=1}^K w_k^2}{w_K^2}.
\]
Under the constrained Gaussian prior $\boldsymbol\theta\sim \mathcal{CN}_K(\boldsymbol 0,\gamma^2\mathbf I_K;\boldsymbol w^\top,0)$,
the normalizing integral over the hyperplane is 
\begin{align}
\label{eq:cn-normalizer}
\int_{\boldsymbol w^\top\boldsymbol\theta=0}
\exp\!\Big\{-\tfrac{1}{2\gamma^2}\boldsymbol\theta^\top\boldsymbol\theta\Big\}\,d\mathcal H^{K-1}(\boldsymbol\theta)
&=\int_{\mathbb R^{K-1}}
\exp\!\Big\{-\tfrac{1}{2\gamma^2}\boldsymbol\theta^{\ast\top}\mathbf B\boldsymbol\theta^\ast\Big\}\,d\boldsymbol\theta^\ast
=\sqrt{\frac{(2\pi\gamma^2)^{K-1}}{\det(\mathbf B)}} .
\end{align}
Using the cluster-collapsed design matrix $\mathbf X_{\boldsymbol z}$, the likelihood \eqref{eq:likelihood} can be expressed as 
\(
\boldsymbol y\mid \boldsymbol\theta,\sigma^2,\boldsymbol z,\mathbf X
~\sim~ \mathcal N(\mathbf X_{\boldsymbol z}\boldsymbol\theta,\ \sigma^2\mathbf I_n).
\)
Therefore, using $\boldsymbol\theta=\mathbf T\boldsymbol\theta^\ast$ and completing the square, we obtain
\begin{align}
&\int \exp\!\Big\{-\tfrac{1}{2\sigma^2}(\boldsymbol y-\mathbf X_{\boldsymbol z}\boldsymbol\theta)^\top
(\boldsymbol y-\mathbf X_{\boldsymbol z}\boldsymbol\theta)\Big\}
\exp\!\Big\{-\tfrac{1}{2\gamma^2}\boldsymbol\theta^\top\boldsymbol\theta\Big\}\,
d\mathcal H^{K-1}(\boldsymbol\theta) \nonumber\\
&\qquad
= \exp\!\Big\{-\tfrac{1}{2\sigma^2}\boldsymbol y^\top\boldsymbol y\Big\}\,
(2\pi\sigma^2)^{\frac{K-1}{2}}\{\det(\mathbf A^\ast)\}^{-\frac12}
\exp\!\Big\{\tfrac{1}{2\sigma^2}\tilde{\boldsymbol b}^{\ast\top}\mathbf A^{\ast-1}\tilde{\boldsymbol b}^\ast\Big\}, \label{eq:num-int}
\end{align}
where 
$
\mathbf A^\ast
=
\mathbf T^\top\mathbf A\mathbf T
$
and 
$
\tilde{\boldsymbol b}^\ast
=
\mathbf T^\top\boldsymbol b.
$
Dividing \eqref{eq:num-int} by the normalizer \eqref{eq:cn-normalizer} and including the 
likelihood constant gives the collapsed marginal likelihood in \eqref{eq:sample_z_fixed} evaluated at the proposed allocation $\boldsymbol z^{*}$,
\begin{align*}
\label{eq:updatey-clean}
f(\boldsymbol y\mid \sigma^2,\gamma^2,\boldsymbol z,\mathbf X)
&= (2 \pi \sigma^2)^{-n/2}
    \bigg(\frac{\sigma^2}{\gamma^2}\bigg)
    ^{\frac{(K-1)}{2}} \{\det (\mathbf{A}^*)\}^{-\frac{1}{2}}\frac{\sqrt{\sum_{k=1}^{K} w_k^2}}{w_K}
    \exp \left\{ -\frac{1}{2\sigma^2} \left( \boldsymbol{y}^\top \boldsymbol{y} - \tilde{\boldsymbol{b}}^{*\top} (\mathbf{A}^*)^{-1}\tilde{\boldsymbol{b}}^*  \right)\right\}.
\end{align*}

\subsubsection{Sampling the cluster parameters \texorpdfstring{$\boldsymbol\theta$}{theta}}

Given the current allocation $\boldsymbol z$, variance parameters
$\sigma^2$ and $\gamma^2$, and response vector $\boldsymbol y$, the full conditional distribution of the cluster effects $\boldsymbol\theta$ is a Gaussian distribution constrained to the hyperplane $\boldsymbol w^\top\boldsymbol\theta=0$. Specifically,
\begin{equation}
\label{eq:theta-cn}
\boldsymbol\theta\mid \boldsymbol z,\sigma^2,\gamma^2,\boldsymbol y
~\sim~ \mathcal{CN}_K\!\big(\boldsymbol\mu_\theta,\mathbf\Sigma_\theta;\,\boldsymbol w^\top,0\big),
\qquad
\mathbf\Sigma_\theta=\Big(\gamma^{-2}\mathbf I_K+\sigma^{-2}\mathbf X_{\boldsymbol z}^\top \mathbf X_{\boldsymbol z}\Big)^{-1},\quad
\boldsymbol\mu_\theta=\mathbf\Sigma_\theta\,\sigma^{-2}\mathbf X_{\boldsymbol z}^\top \boldsymbol y .
\end{equation}
Thus, posterior draws of $\boldsymbol\theta$ satisfy the hyperplane constraint $\boldsymbol w^\top\boldsymbol\theta=0$, which implies that the corresponding regression coefficient vector $\boldsymbol\beta=\mathbf Z\boldsymbol\theta$ satisfies the compositional constraint $\boldsymbol 1^\top\boldsymbol\beta=0$.
To simulate from this constrained Gaussian full conditional, we use the fast algorithm of \citep{cong2017fast}. Specifically, we first draw
$
\boldsymbol\theta^\star \sim N_K(\boldsymbol\mu_\theta,\mathbf\Sigma_\theta),
$
from the unconstrained Gaussian distribution; We then project this draw onto the constraint hyperplane using the covariance-adjusted projection
\[
\boldsymbol\theta
=
\boldsymbol\theta^\star
-
\mathbf\Sigma_\theta \boldsymbol w
\bigl(\boldsymbol w^\top \mathbf\Sigma_\theta \boldsymbol w\bigr)^{-1}
\boldsymbol w^\top \boldsymbol\theta^\star .
\]
The resulting draw satisfies $\boldsymbol w^\top\boldsymbol\theta=0$ and is an exact sample from the constrained Gaussian distribution in \eqref{eq:theta-cn}.

\subsubsection{Sampling the variance parameter \texorpdfstring{$\gamma^2$}{gamma squared}}

Conditional on the current partition, the constrained Gaussian prior for $\boldsymbol\theta$ is supported on a $(K-1)$-dimensional hyperplane. Hence, only $K-1$ degrees of freedom contribute to the update of the slab variance $\gamma^2$. Thus, the full conditional of $\gamma^2$ is given by 
\begin{align}
\label{eq:gamma-update}
p(\gamma^2\mid \boldsymbol\theta,\boldsymbol z,\sigma^2)
~\propto~ (\gamma^2)^{-a_\gamma-\frac{K-1}{2}-1}
\exp\!\left\{-\frac{1}{2\gamma^2}\boldsymbol\theta^\top\boldsymbol\theta
- \frac{b_\gamma}{\gamma^2}\right\}.
\end{align}
Therefore,
\(
\gamma^2\mid\cdot \sim \mathrm{IG}\big(a_\gamma+\tfrac{K-1}{2},\ b_\gamma+\tfrac12\,\boldsymbol\theta^\top\boldsymbol\theta\big).
\)

\noindent \textsl{Edge cases.}
If $K=0$, all predictors are assigned to the spike,  the marginal model reduces to $\boldsymbol y\sim \mathcal N_n(\boldsymbol 0,\sigma^2\mathbf I_n)$ and $\boldsymbol\theta$ is not sampled. $K=1$, the constraint $w_1\theta_1=0$ forces $\theta_1=0$, so this case is equivalent to the all-spike configuration for the purpose of updating $\boldsymbol\theta$ and $\gamma^2$.

\subsubsection{Sampling the variance \texorpdfstring{$\sigma^2$}{sigma squared} and concentration \texorpdfstring{$\alpha$}{alpha}} 
The inverse-gamma prior on $\sigma^2$ leads to a conjugate Gibbs update. Given the current allocation $\boldsymbol z$ and cluster effects $\boldsymbol\theta$, the corresponding full conditional distribution is an inverse gamma density $IG(a_{\sigma} + n/2, b_{\sigma} + 0.5 (\boldsymbol{y - \mathbf{X_{\boldsymbol{z}}\boldsymbol{\theta}}})^\top(\boldsymbol{y - \mathbf{X_{\boldsymbol{z}}\boldsymbol{\theta}}}))$. For the concentration parameter of the Ewens partition prior, we assume
$
\alpha\sim \operatorname{Gamma}(a_\alpha,b_\alpha),
$
and update $\alpha$ using the auxiliary-variable method of
\cite{escobar1995bayesian}. Finally, we set $\alpha_0=2$, which implies $\psi_0\sim \operatorname{Uniform}(0,1)$. This choice gives a diffuse prior on the inclusion probability and allows the data to flexibly determine the number of nonzero coefficients.
 }

\section{Simulation studies}
In this section, we benchmark the performance of BRACE with alternative compositional regression methods \textcolor{blue}{on two simulation scenarios: one with a controlled design that captures structured signal and dependence, and a second which was designed to closely mimic our motivating oral microbiome data set}. 
\label{sec:simulation}
\subsection{General setup}
We start by sampling an $n \times p$ data matrix $\mathbf{U}$ from a multivariate normal distribution $\mathcal{N}_{p}(\boldsymbol{\eta}, \mathbf{\Sigma})$, and obtain the  relative abundance matrix as $\mathbf{O} = \exp(\mathbf{U})/ \boldsymbol{1^\top}\exp(\mathbf{U})$. Using this transformation, the variables follow a logistic normal distribution \citep{aitchison1984log}, a commonly used distribution for modeling microbial abundances. In order to create microbiome features with varying abundance, we set $\eta_j = \log(0.5p)$ for $j = 1,\ldots,10$, and $0$ otherwise, and we assume a covariance structure given by $\mathbf{\Sigma}$. Finally, we generate the responses as $ \boldsymbol{y} = \mathbf{X}\boldsymbol{\beta} + \boldsymbol{\varepsilon}$, where $\boldsymbol{\beta}$ is the vector of regression coefficients and $\mathbf{X} = \log(\mathbf{O})$. We consider settings with $n = 300$ and $p = 100,$ $300$, and $1000$. We generate $\sigma$ so that the signal-to-noise ratio (SNR) is \(\{1,5,10\}\), where $\text{SNR} = \text{mean}(\lvert \boldsymbol{\beta}_{\beta_j \neq 0} \rvert/\sigma, j = 1, \ldots, p)$. For each setting, we generate 100 simulated datasets, and randomly partition the data into training and test samples with a ratio of $80:20$. 

\subsection{Performance metrics}

Each model is fitted on the training set, and the prediction error (PE) $= \frac{1}{n_{\text{test}}} (\boldsymbol{y}_{\text{test}} - \mathbf{X}_{\text{test}}\hat{\boldsymbol{\beta}}_{\text{train}})^{\top}(\boldsymbol{y}_{\text{test}} - \mathbf{X}_{\text{test}}\hat{\boldsymbol{\beta}}_{\text{train}})$ is calculated using the test set,  while the $l_2$ loss $ || \boldsymbol{\beta}_{\text{true}} - \boldsymbol{\hat{\beta}}_{\text{train}}||_{2}$ is calculated using the ground truth coefficient values. {\color{blue}For each test observation \(i\), let \((l_i,u_i)\) denote the \(100(1-\alpha)\%\) posterior predictive interval. The empirical predictive coverage is
\(
\text{Coverage}
=
\frac{1}{n_{\text{test}}}
\sum_{i=1}^{n_{\text{test}}}
1\bigl(y_{\text{test},i} \in (l_i,u_i)\bigr),
\)
and the average predictive interval width is
\(
\text{Width}
=
\frac{1}{n_{\text{test}}}
\sum_{i=1}^{n_{\text{test}}}
(u_i-l_i).
\) We set \(\alpha=0.05\) throughout.}

{\color{blue}\subsection{Variable selection and FDR control}
\label{ssec:VSFDR}
The posterior inclusion probability $\pi_{j1}=\Pr(s_j=1\mid \boldsymbol y)$ provides evidence that feature $j$ belongs to a nonzero cluster, while $\pi_{j0}=\Pr(s_j=0\mid \boldsymbol y)$ captures posterior support for assignment to the point-mass spike.  However, in our model, assignment to an active cluster does not
necessarily imply a practically relevant effect. Because the nonparametric slab can include clusters with atoms close to zero,
a feature may have a relatively high posterior probability of belonging to an active cluster even when its coefficient remains practically indistinguishable from zero.  We therefore use a posterior selection rule that combines evidence of activity with evidence of practical effect size. Specifically, feature $j$ is declared selected if
\[
\Pr(s_j=1 \mid \boldsymbol{y}) > 0.5
\quad \text{and} \quad
\Pr(|\beta_j|>0.1 \mid \boldsymbol{y}) > 0.90.
\]
Equivalently, the rule requires $\pi_{j0}<0.5$ together with $\Pr(|\beta_j|>0.1\mid \boldsymbol y)>0.90$.
This rule has a natural interpretation in terms of a posterior local false discovery rate under a practical null \cite{efron2001empirical, efron2004large}. For a threshold $\varepsilon>0$, define a practical null hypothesis, 
$$
H_{0j}^{(\varepsilon)}:\ |\beta_j|\leq \varepsilon .
$$
Since $\beta_j=0$ whenever $s_j=0$, the posterior local false discovery rate under the practical null can be decomposed as
\[
\mathrm{lfdr}_j^{(\varepsilon)}
=
\Pr\bigl(|\beta_j|\le \varepsilon \,\big|\, \boldsymbol{y}\bigr)
=
\Pr(s_j=0 \mid \boldsymbol{y})
+
\Pr(s_j=1,\ |\beta_j|\leq \varepsilon\mid \boldsymbol y).
\]
The first term is the posterior probability that the feature is in the spike. The second term is the posterior probability that the feature is formally active but assigned to a cluster with negligible effect. Thus, the effect-size condition
\(\Pr(|\beta_j|>\varepsilon \mid \boldsymbol{y}) > 0.90
\)
is equivalent to requiring
\(
\mathrm{lfdr}_j^{(\varepsilon)} < 0.10.
\)
$$
\Pr(|\beta_j|>\varepsilon\mid \boldsymbol y)>0.90
$$
is equivalent to requiring
$$
\mathrm{lfdr}_j^{(\varepsilon)}<0.10.
$$
With $\varepsilon=0.1$, the proposed rule can therefore be viewed as per-feature thresholding of the posterior local false discovery rate at level $q=0.10$. In Supplementary Section S4, we show that this criterion is robust to the choice of $\varepsilon$ and describe the corresponding formal Bayesian FDR step-up procedure for controlling the posterior expected false discovery proportion of the selected set.
 
\subsection{Clustering accuracy} 

In addition to variable selection and prediction, our proposed model enables the grouping of similar features by clustering regression coefficients. In this subsection, we assess the concordance between the true and predicted cluster labels. To obtain the predicted cluster labels, we first post-process the posterior samples to address the label switching problem, a well-documented issue in Bayesian clustering methods, particularly in MCMC-based approaches. Since the likelihood in these models is invariant to permutations of cluster labels, assignments may vary across iterations, even when the underlying clustering structure remains unchanged. This relabeling complicates the interpretation and summarization of posterior samples, necessitating post-processing techniques to ensure label consistency.

To align labels across iterations and derive a stable clustering estimate, we employ the randomized greedy search algorithm SALSO \citep{Dahl2022}. This algorithm selects a point estimate for the cluster labels by optimizing a loss function over posterior MCMC samples. By leveraging SALSO, we obtain a single set of predicted labels that minimizes the posterior expected loss across sampled partitions, ensuring a coherent and interpretable clustering solution.
As recommended by the authors of SALSO, we use the generalized variation of information (VI) loss for the optimization function.  In the simulation studies, clustering accuracy is assessed by comparing the SALSO-estimated labels with the true labels for the nonzero coefficients using the adjusted Rand index (ARI), which has expected value 0 under random label assignment and equals 1 under perfect agreement. We further assess clustering
stability across simulated datasets by computing pairwise ARIs between the SALSO-estimated clusterings from different replicates and reporting the mean cross-replicate ARI.}

We fit BRACE to each simulated training data set using the Gibbs sampler described in Section \ref{sec:sampling}, run for \(8000\) iterations, with the first \(5000\) iterations discarded as burn-in. MCMC chain diagnostics are presented in  Supplementary Section S5. For the primary simulation analyses, we use the prior specifications \(\gamma^2 \sim IG(2.5,1.5)\), \(\sigma^2 \sim IG(0.001,0.001)\), \(a_{\alpha}=1/(0.75\log p)^2\), and \(b_{\alpha}=a_{\alpha}/\sqrt{p}\). This choice of prior for \(\alpha\) allows flexibility in the number of clusters. Further discussion of prior specification for \(\alpha\) is provided in \cite{escobar1995bayesian} and \cite{nott2008predictive}. \textcolor{blue}{To assess sensitivity to the choice of prior on \(\gamma^2\), we additionally carried out a complete sensitivity analysis under \(\gamma^2 \sim IG(5,4)\), with results reported in the Supplementary Material (Section S5, Tables S4 and S5). The main manuscript Tables also present results under both prior configurations for \(\gamma^2\).}

\subsection{Benchmarking methods}
\noindent We compare the performance of BRACE with that of the following existing approaches: \textbf{lasso CLR}, which performs lasso regression\citep{tibshirani1996regression} on centered log ratio (CLR) transformed compositional predictors \citep{aitchison1984log}, \textbf{lasso comp}, a penalized compositional regression approach proposed by \cite{lin2014variable}, \textbf{BAZE}, a Bayesian variable selection algorithm for compositional data\citep{zhang2021bayesian}, that incorporates phylogenetic information using structured Ising priors, and \textbf{BCGLM}, a Bayesian generalized compositional regression model using horseshoe priors \citep{zhang2024bayesian}.  We now describe the parameter settings we used for the benchmarking models. For the Bayesian models, we adopted standard hyperpriors as recommended by the authors. In the case of BCGLM, results remained consistent when varying $m_0$, the prior guess of the number of relevant predictors. Therefore, we set $m_0 = 10$. For BAZE, which relies on a phylogenetic tree-based similarity matrix $\mathbf{Q}$ to capture the similarity between taxa, $\mathbf{Q}$ was defined as a diagonal matrix to reflect a lack of prior information. 
For the frequentist approaches, lasso comp and lasso CLR, cross-validation was used to select the hyperparameters.

\subsection{Simulation settings}
\textcolor{blue}{To evaluate the performance of the proposed approach, we consider two simulation scenarios. Scenario 1 is a controlled setting with prespecified cluster structure and effect sizes, designed to assess the performance of BRACE and competing benchmark methods under controlled conditions similar to those commonly used in the compositional regression literature \cite{lin2014variable,zhang2021bayesian,zhang2024bayesian}. Scenario 2 is a semi-synthetic setting constructed to mimic key characteristics of the ORIGINS oral microbiome study. We next describe these two scenarios and evaluate the performance of BRACE alongside the competing benchmark approaches under each setting.}

\subsection{\textcolor{blue}{Scenario 1: Synthetic data with structured signal and dependence}}
We construct $\boldsymbol{\beta}$ with 35 nonzero elements across 9 clusters, with two clusters having only one element. The true parameter vector is given as $\boldsymbol{\beta} = \{\boldsymbol{-0.8}_{4}, \boldsymbol{-1.41}_{6}, \boldsymbol{-1.95}_{4}, -1.16,
0.96,\boldsymbol{0}_{3}, \boldsymbol{1.04}_{6}, \boldsymbol{0.51}_{4}, \boldsymbol{1.95}_{7}, \boldsymbol{0}_{(p-37)} \}$. It is to be noted that $\boldsymbol{\beta}$ sums to 0 and its components have varying degrees of magnitude, implying varying strengths of association between the  predictors and the response. \textcolor{blue}{This choice was made to reflect several common features of microbiome regression problems: sparsity, since only a subset of taxa is typically associated with the outcome; heterogeneous effect sizes, since associated taxa rarely contribute equally; and clustered signals, to reflect the tendency of biologically related taxa to have similar effects.}

To capture dependence among predictors, we set the within-cluster correlations to ${\Sigma_{i,j}}_{wc}= 0.75 - 0.015\times|i-j|$, making the correlation between two covariates within the same cluster inversely proportional to their distance (with a maximum of 0.75). Additionally, we define the between-cluster correlations among the predictors as ${\Sigma_{i,j}}_{bc}= 0.4 - 0.02\times|i-j|$. The coefficients for the other non-diagonal elements are set to $0$ and the diagonal elements of $\mathbf{\Sigma}$ are set to $1$. \textcolor{blue}{This covariance structure was chosen to mimic the stronger correlation typically observed among related taxa than among unrelated taxa, while still allowing moderate between-group dependence. Overall, Scenario 1 provides a controlled yet biologically motivated setting for evaluating whether BRACE can recover sparse and clustered effects under realistic predictor dependence.}

Table \ref{tab:sim_res2} presents the results for Scenario 1, comparing BRACE against six competing methods. BRACE consistently achieves the lowest or near-lowest prediction error and L2 loss across nearly all settings, with its best entries marked in bold. The Bayesian competitors (BAZE, BCGLM) also perform well on PE and L2 loss relative to the lasso variants, but BRACE tends to edge them out, particularly in higher dimensions. Notably, BCGLM shows an exceptionally high computation time for $p=1000$  which limits the results to 10 replicates. 

\color{blue}
\textbf{Variable selection and FDR control.}
Scenario 1, with its distinct within- and between-cluster covariance structures, presents a challenging environment for variable selection. Under the baseline $\pi_j > 0.5$ criterion, BRACE recovers all 34 active variables at $p = 300$, but the near-zero cluster phenomenon is substantially pronounced: the false discovery proportion (FDP) reaches values close to 0.5 at $p = 300$ (Supplementary Table S4). At $p = 100$, under the baseline rule, the FDP remains elevated at around 0.2.
Augmenting the PIP threshold with the probabilistic magnitude filter
\(
P(|\beta_j| > \varepsilon \mid \boldsymbol{y}) > 0.90
\)
eliminates false positives entirely across all priors, dimensionalities, and values of $\varepsilon \in \{0.05, 0.10, 0.20\}$, with the FDP falling to zero and no loss in true positive detection at $p = 300$ (Supplementary Table S5).  These results reinforce the importance of the practical-null criterion in correlated designs: the severity of FDR inflation under the baseline rule grows with the complexity of the dependence structure, yet the combined rule consistently restores control. The variable selection results reported in Table~\ref{tab:sim_res2} correspond to the practical thresholding rule with $\varepsilon = 0.1$.
\color{black}

\textbf{Clustering accuracy.}
\textcolor{blue}{
Table~\ref{tab:brace_combined_sim_real} (Panel A) summarizes BRACE's cluster recovery for Scenario 1. At SNR $= 1$, the adjusted Rand index ranges from $0.93$ to $0.95$ across $p \in \{100, 300, 1000\}$, indicating that BRACE accurately recovers the true grouped coefficient structure even under low signal. Cluster recovery improves with signal strength, reaching $\text{ARI} = 1.00$ at SNR $\ge 5$ for $p \ge 300$, with the correct number of clusters ($8$) consistently identified. At $p = 100$, ARI remains high ($0.99$) at SNR $= 5$ and $10$, although cross-replicate ARI values of $0.63$--$0.70$ indicate some variability in exact cluster boundaries across independent datasets, a consequence of the smaller null class providing less contrast for cluster resolution. For $p \ge 300$, cross-replicate ARI reaches $1.00$ at SNR $\ge 5$, confirming that the inferred clustering is fully stable across data realizations.}

{\color{blue}BRACE also provides well-calibrated uncertainty quantification. Predictive credible interval coverage remains at or near the nominal $95\%$ level across all settings, ranging from $0.92$ at $(p = 1000, \text{SNR} = 1)$ to $0.99$ at $(p = 300, \text{SNR} = 10)$. Interval widths narrow appropriately with increasing signal, from approximately $5.2$--$5.8$ at SNR $= 1$ to $0.5$--$0.7$ at SNR $= 10$, reflecting proper posterior concentration. Notably, coverage is maintained even as intervals tighten, indicating that the posterior uncertainty is well calibrated.}


Overall, these results highlight the strong performance of BRACE in this challenging setting and underscore its relevance for microbiome applications, where many rare features may share similar effects and benefit from joint modeling through clustering. BRACE consistently delivers competitive or superior out-of-sample prediction accuracy while maintaining strong control of false positives relative to the competing methods in this controlled simulation setting. In addition, the uniformly high ARI values across all scenarios demonstrate that BRACE is able to recover the true cluster structure with high accuracy.

We considered this simulation setting as a benchmark study for comparing the computational cost of the methods considered. We found that for the setting with $p=100$, BCGLM required the longest run times (95 seconds for 1000 MCMC iterations), followed by BRACE (83 seconds for 1000 MCMC iterations) on a MacBook Pro with 16GB RAM. The penalized methods and BAZE were more computationally efficient. 

We further considered two additional simulation settings. In the first, we retained the coefficient structure of Scenario 1 but imposed a simpler autoregressive correlation structure on the predictors, with \(\Sigma_{ij}=\rho^{|i-j|}\), \(\rho=0.5\), and \(\mathrm{SNR}=1\); the results are reported in Supplementary Table~S6. In the second, we examined a setting in which the true regression coefficient vector \(\boldsymbol{\beta}\) contained no cluster structure; details are provided in Supplementary Section S5.3, with results reported in Supplementary Table~S7. In both settings, BRACE maintained superior performance relative to the competing methods across \(p \in \{100,300,1000\}\).

\begin{table}[h!]
\caption{\textcolor{blue}{Performance comparison for Scenario 1 with dependent covariates with different between and within cluster covariance setup, \textbf{sample size $n=300$}, and \textbf{$\text{SNR} = 1$}. Revised Performance is summarized in terms of prediction error (PE), L2 loss in estimation of the coefficient vector, and number of false positives (FP) and false negatives (FN). Entries that reflect the smallest PE and L2 loss are marked in bold.}} 
\begin{center}
\resizebox{0.75\textwidth}{!}{
\begin{tabular}{ c c c c c c } 
\hline
& Method & PE & L2 Loss & FP& FN  \\
\hline
p = 100 \& & lasso CLR & 12472 (1804.1) & 0.39 (0.05) & 0.04 (0.02) & 0.00 (0.00)\\ 
SNR = 1 &lasso constrained & 3918.1 (908.0)  & 8.43 (0.29) & 0.04 (0.2) & 32.64 (1.12)\\
&lasso comp & 4.02 (1.04)  & 0.43 (0.07)  & 0.80 (0.95) & 0.00 (0.00)\\  
&BAZE & 2.16 (0.19)  & 0.24 (0.03) &  0.00 (0.00) & 0.00 (0.00)\\ 
&BCGLM &  2.47 (0.34)  & 0.20 (0.01)  & 1.64 (1.52) & 0.00 (0.00)\\ 

& {\color{blue} \textbf{BRACE}($\gamma^2 \sim IG(5,4)$)} & {\color{blue} 1.83 (0.14)} & {\color{blue} 0.11 (0.03)} & {\color{blue} 0.00 (0.00)} & {\color{blue} 0.00 (0.00)} \\
\hline
p = 300 \& & lasso CLR & 17715 (2110.5) & 0.48 (0.06) & 0.05 (0.22) &  0.00 (0.00)\\ 
 SNR = 1 &lasso constrained & 3989.8 (754.0) & 8.62(0.56) & 0.15 (0.36) & 32.64(1.15)\\ 
&lasso comp & 1360.3 (283.8)  & 6.40 (0.25) & 0.00 (0.00) & 16.88 (1.46)\\ 
&BAZE & 2.59 (0.19)  & 0.42 (0.05) &  0.00 (0.00) & 0.00 (0.00)\\ 
&BCGLM & 2.47 (0.34)  & 0.21 (0.01)  & 2.40 (2.56) & 0.00 (0.00)\\ &{\color{blue} \textbf{BRACE}} & {\color{blue} \textbf{1.75} (0.05)} & {\color{blue} \textbf{0.09} (0.03)} & {\color{blue} 0.00(0.00)} & {\color{blue} 0.00(0.00)} \\
& {\color{blue} \textbf{BRACE}($\gamma^2 \sim IG(5,4)$)} & {\color{blue} 2.05 (0.13)} & {\color{blue} 0.09 (0.02)} & {\color{blue} 0.00 (0.00)} & {\color{blue} 0.00 (0.00)} \\

\hline
p = 1000 \& & lasso CLR & 25016.4 (2782.8)  & 0.49 (0.05) & 0.2 (0.53) & 0.00 (0.00)\\ 
SNR = 1 &lasso constrained & 4216.4 (612.7) & 8.81 (0.86) & 0.05 (0.23) & 32.89 (1.08) \\ 
&lasso comp & 1491.1 (334.6) & 6.52 (0.31) & 0.00 (0.00) & 17.35 (1.66)\\ 
&BAZE & 2.23 (0.19)  & 0.17 (0.03) & 0.00 (0.00)  & 0.00 (0.00)\\ 
&BCGLM & 21.756 (17.0)  & 0.11 (0.03) & 0.00 (0.00)& 12.5 (3.53)\\ 
&{\color{blue} \textbf{BRACE}} & {\color{blue} \textbf{2.36} (0.12)} & {\color{blue} 0.10 (0.01)} & {\color{blue} 0.00 (0.00)} & {\color{blue} 0.00 (0.00)} \\ 
& {\color{blue} \textbf{BRACE}($\gamma^2 \sim IG(5,4)$)} & {\color{blue} 2.41 (0.11)} & {\color{blue} 0.10 (0.02)} & {\color{blue} 0.00 (0.00)} & {\color{blue} 0.00 (0.00)} \\
\hline
\hline 
p = 100 \& & lasso CLR & 12439.76 (2660.68) & 0.40 (0.06) & 0.00 (0.00) & 0.00 (0.00)\\ 
SNR = 5 &lasso constrained & 3584.45(116.16)  & 8.39 (0.08) & 0.00 (0.00) & 32.50 (2.00)\\
&lasso comp & 2.17 (0.70)  & 0.38 (0.05)  & 0.00 (0.00) & 0.00 (0.00)\\  
&BAZE & 0.12 (0.01)  & 0.01 (0.01) &  0.00 (0.00) & 0.00 (0.00)\\ 
&BCGLM &  0.14 (0.01)& 0.01 (0.01) &1.00 (1.00)   &0.00 (0.00)\\ 
& {\color{blue} \textbf{BRACE}} & {\color{blue} \textbf{0.11} (0.02)} & {\color{blue} \textbf{0.01} (0.01)} & {\color{blue} 0.00 (0.00)} & {\color{blue} 0.00 (0.00)}\\ 
& {\color{blue} \textbf{BRACE}($\gamma^2 \sim IG(5,4)$)} & {\color{blue} 0.10 (0.01)} & {\color{blue} 0.01 (0.00)} & {\color{blue} 0.00 (0.00)} & {\color{blue} 0.00 (0.00)} \\
\hline
p = 300 \& & lasso CLR & 17153.43 (3611.36) &0.45(0.06) & 0.00 (0.00) &  0.00 (0.00)\\ 
 SNR = 5 &lasso constrained & 3875.10 (928.10) & 8.39(0.15) & 0.00 (0.00) & 32.00(2.00)\\ 
&lasso comp & 1298.23 (387.22)  & 6.34 (0.37) & 0.00 (0.00) & 17.00 (2.00)\\ 
&BAZE & 0.10 (0.01)  & 0.04 (0.01) &  0.00 (0.00) & 0.00 (0.00)\\ 
&BCGLM & 0.20 (0.03) & 0.01 (0.01) & 2.00 (2.00) &  0.00 (0.00)\\ 
&{\color{blue} \textbf{BRACE}} & {\color{blue} \textbf{0.10} (0.02)} & {\color{blue} \textbf{0.01} (0.01)} & {\color{blue} 0.00(0.00)} & {\color{blue} 0.00(0.00)} \\
& {\color{blue} \textbf{BRACE}($\gamma^2 \sim IG(5,4)$)} & {\color{blue} 0.09 (0.01)} & {\color{blue} 0.01 (0.00)} & {\color{blue} 0.00 (0.00)} & {\color{blue} 0.00 (0.00)} \\
\hline
p = 1000 \& & lasso CLR & 24939.51 (3819.03)  & 0.44 (0.05) & 0.2 (0.53) & 0.00 (0.00)\\ 
SNR = 5 &lasso constrained & 3829.05 (1025.25) &8.41 (0.19) & 0.00 (0.00) & 32.00 (2.00) \\ 
&lasso comp & 1523.62 (608.03) & 6.43 (0.37) & 0.00 (0.00) & 18.00 (2.00)\\ 
&BAZE & 0.09 (0.01)  & 0.04 (0.01) & 0.00 (0.00)  & 0.00 (0.00)\\ 
&BCGLM & 13.89 (5.47) & 0.01 (0.01) &0.00 (0.00) &1.00 (0.50)\\ 
&{\color{blue} \textbf{BRACE}} & {\color{blue} \textbf{0.08} (0.01)} & {\color{blue} 0.01(0.00)} & {\color{blue} 0.00 (0.00)} & {\color{blue} 0.00 (0.00)}\\ 
&{\color{blue} \textbf{BRACE}($\gamma^2 \sim IG(5,4)$)} & {\color{blue} \textbf{0.07} (0.01)} & {\color{blue} 0.01(0.00)} & {\color{blue} 0.00 (0.00)} & {\color{blue} 0.00 (0.00)}\\ 
\hline
\hline
p = 100 \& & lasso CLR & 12925.463 (1480.53) & 0.42 (0.06) & 0.00 (0.00) & 0.00 (0.00)\\ 
SNR = 10 &lasso constrained & 3881.90 (512.85)  & 8.39 (0.13) & 0.00 (0.0) & 32.50 (2.00)\\
&lasso comp & 2.30 (0.70)  & 0.37 (0.06)  & 0.00 (0.00) & 0.00 (0.00)\\  
&BAZE & 0.03 (0.01)  & 0.00 (0.00) &  0.00 (0.00) & 0.00 (0.00)\\ 
&BCGLM &  0.03 (0.00) & 0.00 (0.00)   &1.00 (1.00)   &  0.00 (0.00)\\ 
& {\color{blue} \textbf{BRACE}} & {\color{blue} \textbf{0.02} (0.01)} & {\color{blue} \textbf{0.00} (0.00)} & {\color{blue} 0.00 (0.00)} & {\color{blue} 0.00 (0.00)}\\ 
& {\color{blue} \textbf{BRACE}($\gamma^2 \sim IG(5,4)$)} & {\color{blue} 0.02 (0.00)} & {\color{blue} 0.01 (0.00)} & {\color{blue} 0.00 (0.00)} & {\color{blue} 0.00 (0.00)} \\
\hline
p = 300 \& & lasso CLR & 17925.46 (846.57) & 0.46 (0.20) & 0.05 (0.22) &  0.00 (0.00)\\ 
 SNR = 10 &lasso constrained & 3800.48 (846.57) & 8.39 (0.21) & 0.00 (0.00) & 33.00 (2.00)\\ 
&lasso comp & 1368.25 (306.4) & 6.37 (0.42)& 0.00 (0.00) & 18.00 (2.00)\\ 
&BAZE & 0.02 (0.01)  & 0.02 (0.00) &  0.00 (0.00) & 0.00 (0.00)\\ 
&BCGLM & 0.05 (0.01) & 0.00 (0.00) & 3.00 (0.50)   &  0.00 (0.00)\\ 
&{\color{blue} \textbf{BRACE}} & {\color{blue} \textbf{0.02} (0.01)} & {\color{blue} \textbf{0.00} (0.00)} & {\color{blue} 0.00(0.00)} & {\color{blue} 0.00(0.00)} \\
& {\color{blue} \textbf{BRACE}($\gamma^2 \sim IG(5,4)$)} & {\color{blue} 0.01 (0.00)} & {\color{blue} 0.01 (0.00)} & {\color{blue} 0.00 (0.00)} & {\color{blue} 0.00 (0.00)}\\
\hline
p = 1000 \& & lasso CLR & 24911.48 (3139.65) &0.46 (0.07) &0.00 (0.00)  & 0.00 (0.00)\\ 
SNR = 10 &lasso constrained & 4403.06  (896.69)& 8.39 (0.12) & 0.00 (0.00) & 32.00 (2.00)\\ 
&lasso comp & 1547.00 (255.53) &6.37 (0.42)&  0.00 (0.00) &18.00 (2.00)\\ 
&BAZE & 0.02 (0.01)  & 0.02 (0.01) & 0.00 (0.00)  & 0.00 (0.00)\\ 
&BCGLM & 3.40 (0.66) & 0.00 (0.00)   & 0.00 (0.00) & 0.50 (0.50)\\ 
&{\color{blue} \textbf{BRACE}} & {\color{blue} \textbf{0.02} (0.01)} & {\color{blue} 0.00(0.00)} & {\color{blue} 0.00 (0.00)} & {\color{blue} 0.00 (0.00)}\\ 
& {\color{blue} \textbf{BRACE}($\gamma^2 \sim IG(5,4)$)} & {\color{blue} 0.02 (0.00)} & {\color{blue} 0.01 (0.00)} & {\color{blue} 0.00 (0.00)} & {\color{blue} 0.00 (0.00)} \\
\hline
\end{tabular}}
\end{center} \label{tab:sim_res2}
\end{table}


{\color{blue}\subsection{Scenario 2: Semi-synthetic data mimicking the ORIGINS dataset}

We conducted a semi-synthetic simulation study to evaluate method performance under realistic conditions. Rather than generating entirely artificial data, we preserved the complex correlation structure inherent in real microbiome data by using covariate matrices derived from the ORIGINS study. The processed dataset comprised $n = 111$ participants and $p = 130$ bacterial taxa at the species level.

Prior to simulation, the taxa abundance table underwent the following preprocessing steps. To enable log-transformation of compositional data, zero counts were replaced by adding half the minimum observed non-zero abundance to all zero entries:
this multiplicative replacement strategy preserves the compositional nature of the data while avoiding arbitrary pseudocount selection. Subsequently, the abundance matrix was converted to relative abundances by dividing each entry by its corresponding sample total, yielding the compositional matrix $\tilde{\mathbf{U}}\in \mathbb{R}^{n \times p}$ where $\sum_{j=1}^{p} \tilde{u}_{ij} = 1$ for all $i$.

To generate regression coefficients that reflect realistic effect patterns observed in microbiome studies, we employed a data-driven approach. First, compositional lasso regression \citep{lin2014variable} was applied to the real data with fasting insulin levels as the response variable to obtain preliminary coefficient estimates respecting the compositional constraint. Second, $k$-means clustering with $k = 6$ groups was applied to the estimated coefficients to identify natural groupings, and coefficients with absolute values below 0.1 were set to zero to induce sparsity. This reflects the biological expectation that only a subset of taxa are truly associated with the outcome. Third, non-zero coefficients were centered to satisfy the sum-to-zero constraint required for compositional regression:
\(
    \sum_{j \in \mathcal{S}} \beta_j = 0,  \text{where } \mathcal{S} = \{j : \beta_j \neq 0\}
\).
Finally, the coefficient vector was scaled by a factor of 2 to ensure adequate signal strength across the range of signal-to-noise ratios examined. The true coefficient vector $\boldsymbol{\beta}$ was sparse, with 22 of 130 coefficients (16.9\%) being non-zero. The non-zero coefficients exhibited a grouped structure arising from the $k$-means clustering procedure, taking values from the set:
\(
    \{1.19_{\times 9}, \, -0.39_{\times 6}, \, -1.09_{\times 4}, \, -0.82_{\times 2}, \, -2.39_{\times 1}\}
\)
where subscripts denote multiplicities. This configuration reflects both the sparsity typical of microbiome-outcome associations and the biological plausibility that taxonomically or functionally related taxa may share similar effect magnitudes.

Bootstrap samples of size $n = 111$ were drawn with replacement from the rows of the real relative abundance matrix, preserving the empirical correlation structure among taxa. Let $\tilde{\mathbf{U} }\in \mathbb{R}^{n \times p}$ denote the resampled compositional matrix. The log-transformed design matrix was computed as $\mathbf{X} = \log(\tilde{\mathbf{U}})$. Continuous response variables were then generated according to the linear model:
\(
    \mathbf{y} = \mathbf{X} \boldsymbol{\beta} + \boldsymbol{\varepsilon}, \quad \boldsymbol{\varepsilon} \sim \mathcal{N}(\mathbf{0}, \sigma^2 \mathbf{I}_n)
\)
where $\boldsymbol{\beta} \in \mathbb{R}^p$ denotes the true sparse coefficient vector satisfying the compositional constraint. 

To assess method performance across varying noise levels, we considered three signal-to-noise ratio (SNR) settings: $\text{SNR} \in \{1, 5, 10\}$. The noise standard deviation $\sigma$ was calibrated according to:
\(
    \sigma = \frac{\bar{|\beta|}}{\text{SNR}}, \quad \text{where } \bar{|\beta|} = \frac{1}{|\mathcal{S}|} \sum_{j \in \mathcal{S}} |\beta_j|
\)
and $\mathcal{S} = \{j : \beta_j \neq 0\}$ denotes the active set. This formulation ensures that $\sigma$ scales with the average magnitude of true effects, providing interpretable SNR levels: $\text{SNR} = 1$ represents a low-signal setting where noise magnitude equals the average effect size, while $\text{SNR} = 10$ represents a high-signal setting with substantially lower noise. For each SNR setting, 30 independent replicates were generated to assess variability in method performance.

Table \ref{tab:fixednp_varysnr} reports results across three signal-to-noise ratios. BRACE achieves the lowest prediction error and L2 loss across all SNR levels. At SNR = 1, BRACE attains a median PE of 1.16 compared to 4.65 for the next best method (BCGLM), with the gap widening at higher signal: PE reduces to 0.08 and 0.01 at SNR = 5 and 10, while no competitor falls below 1.59. For variable selection at SNR = 1, BRACE achieves the best balance of false positives (2.50) and false negatives (2.80), whereas BAZE and BCGLM produce near-zero false positives but miss the majority of true signals (21.07 and 15.00 false negatives), and lasso CLR exhibits 34.00 false positives. At SNR $\ge$ 5, BRACE achieves perfect selection with zero false positives and zero false negatives, a property not shared by any competing method. Results under the alternative prior $\gamma^2 \sim IG(5,4)$ are nearly identical, confirming robustness.

Table \ref{tab:brace_combined_sim_real} (Panel B) provides additional diagnostics for BRACE. Predictive credible intervals maintain nominal or near-nominal coverage across all SNR levels (0.95, 0.92, and 0.97), with widths narrowing from 4.26 at SNR = 1 to 0.40 at SNR = 10, reflecting appropriate posterior concentration as signal increases. Clustering recovery, measured by the adjusted Rand index, improves from 0.72 at SNR = 1 to 1.00 at SNR $\geq$ 5, indicating perfect recovery of the true grouped coefficient structure once sufficient signal is available. Cross-replicate ARI similarly reaches 1.00 at SNR $\ge$ 5, confirming that the clustering solution is stable across independent data realizations, while the lower value of 0.65 at SNR = 1 appropriately reflects greater uncertainty in the low-signal regime.

\begin{table}[htbp]
\centering
\caption{Performance comparison for Scenario 2 with fixed $n= 111$, $p=130$, and varying SNR. 
Entries are reported as median (IQR) across 30 replicates.}
\label{tab:fixednp_varysnr}
\setlength{\tabcolsep}{8pt}
\renewcommand{\arraystretch}{1.15}

\begin{tabular}{>{\raggedright\arraybackslash}p{2cm} l c c c c }
\toprule
 & \textbf{Method} & \textbf{PE} & \textbf{L2 Loss} & \textbf{FP} & \textbf{FN}  \\
\midrule

\multirow{6}{*}{\begin{tabular}[c]{@{}l@{}}SNR $=1$\end{tabular}}
& lasso CLR         &33.94 (17.45) & 2.30 (0.86) & 34.00 (5.75) & 2.50 (2.75)   \\
& lasso constrained & 26.45 (12.81) & 5.04 (0.03) &  1.50 (2.00)  &   22.00 (0.00)   \\
& lasso comp        &  20.69 (9.22) & 4.58 (0.36)  &   6.00 (3.00) &  18.00 (2.75)   \\
& BAZE              & 6.94 (11.35)  & 4.88(0.35) & 0.03 (0.18) & 21.07 (1.64)   \\
& BCGLM             &  4.65 (3.29)& 0.06 (0.02) &0.00 (0.00)   &15.00 (2.00)   \\
& \textcolor{blue}{\textbf{BRACE}}
                    & 1.16 (0.37)& 0.75 (0.19) & 2.50(2.00) & 2.80 (2.00) \\
& \textcolor{blue}{\textbf{BRACE}} $(\gamma^2 \sim \mathrm{IG}(5,4))$
                    & 1.14 (0.35) & 0.73 (0.16) & 2.00 (2.00) & 1.00 (1.00) \\
\midrule

\multirow{6}{*}{\begin{tabular}[c]{@{}l@{}}SNR $=5$\end{tabular}}
& lasso CLR         &  40.67 (17.01) & 1.15 (0.69) & 27.00 (6.75) & 1.00 (2.00)    \\
& lasso constrained &  27.84 (10.52) & 5.05 (0.03)   &  2.00 (2.00) & 22.00 (0.00)    \\
& lasso comp        &  21.77 (8.58) & 4.65 (0.23)  &    6.00(3.00) & 17.00 (2.00)    \\
& BAZE              & 3.18 (7.70) &5.02 (0.28)& 0.10 (0.40) & 21.47 (1.36)   \\
& BCGLM             &  2.31 (5.34) &0.05 (0.02)& 0.00 (0.00) & 14.00 (3.50)    \\
& \textcolor{blue}{\textbf{BRACE}}
                    & 0.08 (0.01) & 0.06(0.01) & 0.00 (0.00) & 0.00 (0.00) \\
 & \textcolor{blue}{\textbf{BRACE}} $(\gamma^2 \sim \mathrm{IG}(5,4))$
                    & 0.08 (0.01) & 0.05 (0.03) & 0.00 (0.00) & 0.00 (0.00) \\
\midrule

\multirow{6}{*}{\begin{tabular}[c]{@{}l@{}}SNR $=10$\end{tabular}}
& lasso CLR         &  36.3 (24.06) & 1.07 (0.78)  &  27.00 (7.00) & 1.00 (2.00)    \\
& lasso constrained &  27.33 (13.34) & 5.05 (0.02)  &   2.00 (2.00) & 22.00 (0.00)   \\
& lasso comp        &  23.05 (10.77) & 4.67 (0.27) & 6.50 (2.50) & 18.00 (1.00)   \\
& BAZE              &  3.41 (7.92)& 4.92 (0.38)  & 0.07 (0.25) & 21.23 (2.06)   \\
& BCGLM             &  1.59 (1.17)&    0.04 (0.00) & 0.00 (0).00   & 11.00 (1.00)    \\
& \textcolor{blue}{\textbf{BRACE}}
                    & 0.01 (0.00)  & 0.02 (0.01) & 0.00 (0.00) &  0.00 (0.00)\\
& \textcolor{blue}{\textbf{BRACE}} $(\gamma^2 \sim \mathrm{IG}(5,4))$
                    & 0.01 (0.00) & 0.02 (0.01) & 0.00 (0.00) & 0.00 (0.00) \\
\bottomrule
\end{tabular}
\end{table}
}


\begin{table}[!ht]
\centering
\scriptsize
\caption{\textcolor{blue}{Summary of BRACE performance for Scenarios 1 and 2. For Scenario 1 (Panel A), results are reported across $p \in \{100,300,1000\}$ and $\mathrm{SNR} \in \{1,5,10\}$. For Scenario 2 (Panel B), results are shown for $p=130$, $n=111$, and $\mathrm{SNR} \in \{1,5,10\}$. Entries are reported as mean (SD) across replicates and summarize predictive coverage, interval width, clustering recovery, cross-replicate clustering stability, and the number of inferred clusters.}}
\label{tab:brace_combined_sim_real}
\setlength{\tabcolsep}{6pt}
\color{blue}
\begin{tabular}{ccccccc}
\toprule
$p$ & SNR & Predictive CI & Predictive Width & ARI & Cross-Replicate-ARI & \# Clusters \\
\midrule
\multicolumn{7}{l}{\textbf{Panel A: Scenario 1}} \\
\midrule
100  & 1  & 0.96 (0.02) & 5.83 (0.06) & 0.93 (0.03) & 0.67 (0.39) & 8 (2) \\
100  & 5  & 0.90 (0.04) & 1.04 (0.01) & 0.99 (0.01) & 0.63 (0.47) & 8 (1) \\
100  & 10 & 0.95 (0.02) & 0.52 (0.01) & 0.99 (0.00) & 0.70 (0.45) & 8 (1) \\
\addlinespace
300  & 1  & 0.93 (0.02) & 5.17 (0.06) & 0.95 (0.03) & 0.94 (0.05) & 7 (2) \\
300  & 5  & 0.94 (0.01) & 1.15 (0.01) & 1.00 (0.00) & 1.00 (0.00) & 8 (1) \\
300  & 10 & 0.99 (0.01) & 0.56 (0.00) & 1.00 (0.00) & 1.00 (0.00) & 8 (1) \\
\addlinespace
1000 & 1  & 0.92 (0.02) & 5.84 (0.06) & 0.94 (0.02) & 0.96 (0.04) & 7 (2) \\
1000 & 5  & 0.94 (0.01) & 1.10 (0.01) & 1.00 (0.00) & 1.00 (0.00) & 8 (1) \\
1000 & 10 & 0.97 (0.01) & 0.69 (0.00) & 1.00 (0.00) & 1.00 (0.00) & 8 (1) \\

\midrule
\multicolumn{7}{l}{\textbf{Panel B: Scenario 2 ($p=130$, $n=111$)}} \\
\midrule
130  & 1  & 0.95 (0.05) & 4.26 (0.32) & 0.72 (0.11) & 0.65 (0.20) & 9 (1) \\
130  & 5  & 0.92 (0.04) & 1.08 (0.01) & 1.00 (0.00) & 1.00 (0.00) & 5 (1) \\
130  & 10 & 0.97 (0.04) & 0.40 (0.01) & 1.00 (0.01) & 1.00 (0.01) & 5 (0) \\

\addlinespace
\multicolumn{7}{l}{\textit{Additional BRACE results for}  $\gamma^2 \sim \mathrm{IG}(5,4)$} \\
\midrule
\multicolumn{7}{l}{\textbf{Panel A: Scenario 1}} \\
\midrule
100 & 1  & 0.96 (0.01) & 5.84 (0.06)  & 0.93 (0.03) & 0.85 (0.22) & 8 (2) \\
100 & 5  & 0.89 (0.03) & 1.04 (0.01)  & 0.83 (0.37) & 0.69 (0.45) & 8 (0) \\
100 & 10 & 0.95 (0.03) & 0.54 (0.01) & 0.74 (0.43) & 0.55 (0.48) & 8 (0) \\
\addlinespace
300 & 1  & 0.93 (0.02) & 5.17 (0.05)  & 0.96 (0.01) & 0.94 (0.05) & 7 (1) \\
300  & 5  & 0.95 (0.01) & 1.15 (0.01) & 1.00 (0.00) & 1.00 (0.00) & 8 (1) \\
300  & 10 & 0.99 (0.01) & 0.56 (0.00) & 1.00 (0.00) & 1.00 (0.00) & 8 (1) \\
\addlinespace
1000 & 1  & 0.92 (0.02) & 5.84 (0.06) & 0.94 (0.02) & 0.96 (0.04) & 7 (2) \\
1000 & 5  & 0.94 (0.01) & 1.10 (0.01) & 1.00 (0.00) & 1.00 (0.00) & 8 (1) \\
1000 & 10 & 0.97 (0.01) & 0.71 (0.00) & 1.00 (0.00) & 1.00 (0.00) & 8 (1) \\
\midrule
\multicolumn{7}{l}{\textbf{Panel B: Scenario 2 ($p=130$, $n=111$)}} \\
\midrule
130 & 1  & 0.95 (0.04) & 4.25 (0.29) & 0.73 (0.12) & 0.70 (0.19) & 9 (0) \\
130 & 5  & 0.85 (0.03) & 0.76 (0.02) & 0.97 (0.06) & 0.95 (0.08) & 5 (0) \\
130 & 10 & 0.97 (0.03) & 0.41 (0.01) & 1.00 (0.00) & 1.00 (0.00) & 5 (0) \\

\bottomrule
\end{tabular}
\end{table}

\section{\textcolor{blue}{Real data applications}}
\label{sec:case_study}
To illustrate the utility of our proposed method, we applied it to oral microbiome data from the Oral Infections, Glucose Intolerance, and Insulin Resistance Study (ORIGINS), which investigated the correlation between periodontal microbiota and insulin resistance \citep{demmer2017subgingival}. Previous studies have established a significant association between periodontitis, a chronic inflammatory disease affecting the tissues supporting the teeth, and the risk of type 2 diabetes. Type 2 diabetes, constituting 90\% of diabetes cases, arises from disruptions in glucose regulation and insulin resistance. The cross-sectional ORIGINS study included 152 adults without diabetes (77\% female), aged 20–55 years. 
The  Human Oral Microbe Identification Microarray \citep{colombo2009} was used to quantify the abundance of 379 taxa in subgingival plaque samples. For this case study, we obtained the microbiome profiling data from 
 \cite{demmer2017subgingival}. We utilized the observed fasting insulin levels as our response variable, employing our proposed method to elucidate the relationship between the periodontal microbiome and insulin levels. \textcolor{blue}{We filtered the dataset to exclude repeated samples, samples missing insulin level information, and taxa with prevalence below 1\%, resulting in 130 taxa and 111 samples.} After filtering, there were no repeated measures (i.e., each sample corresponds to a unique subject). \textcolor{blue}{Zero counts in the taxa count matrix were addressed using pseudocount-based preprocessing, which is commonly used in microbiome analyses \citep{lin2014variable, bien2021tree, mishra2024taro}. Each zero entry was replaced by 
0.5 $\times$ the minimum observed nonzero abundance. Relative species compositions were then computed, log-transformed, and used as inputs to the regression model. To assess the robustness of our results to zero handling via pseudocounts, we repeated the analysis under an alternative pre-processing scheme \cite{shi2022high}. Full details are provided in Section S6.1 of the Supplementary Material.}

\subsection{Prediction and selection results}
\label{subsec::Prediction and selection results}
\begin{figure}[htbp]
    \centering
    
    \includegraphics[
        width=0.95\textwidth,
        trim={0cm 5cm 0cm 4cm},
        clip
    ]{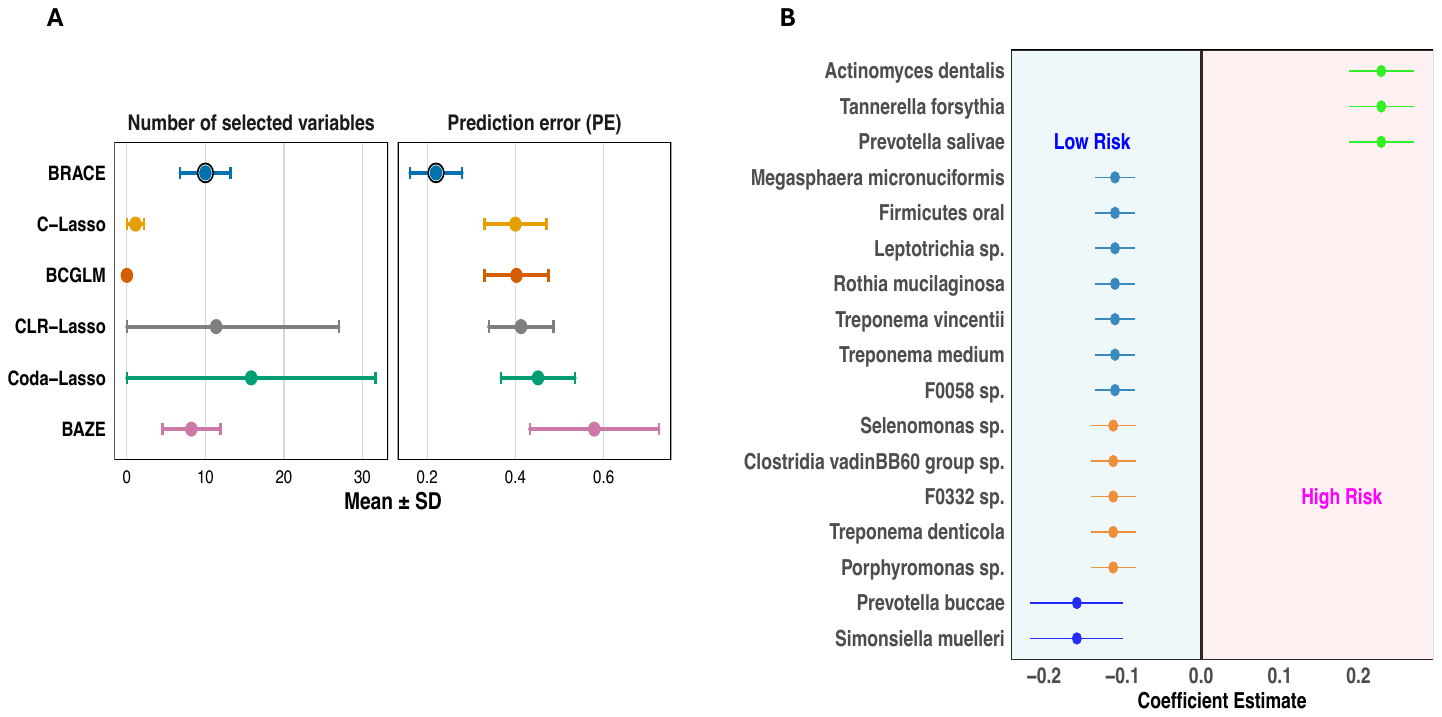}
    
    \vspace{-0.4cm}
    
    \includegraphics[
        width=0.95\textwidth,
        trim={0cm 2cm 0cm 3cm},
        clip
    ]{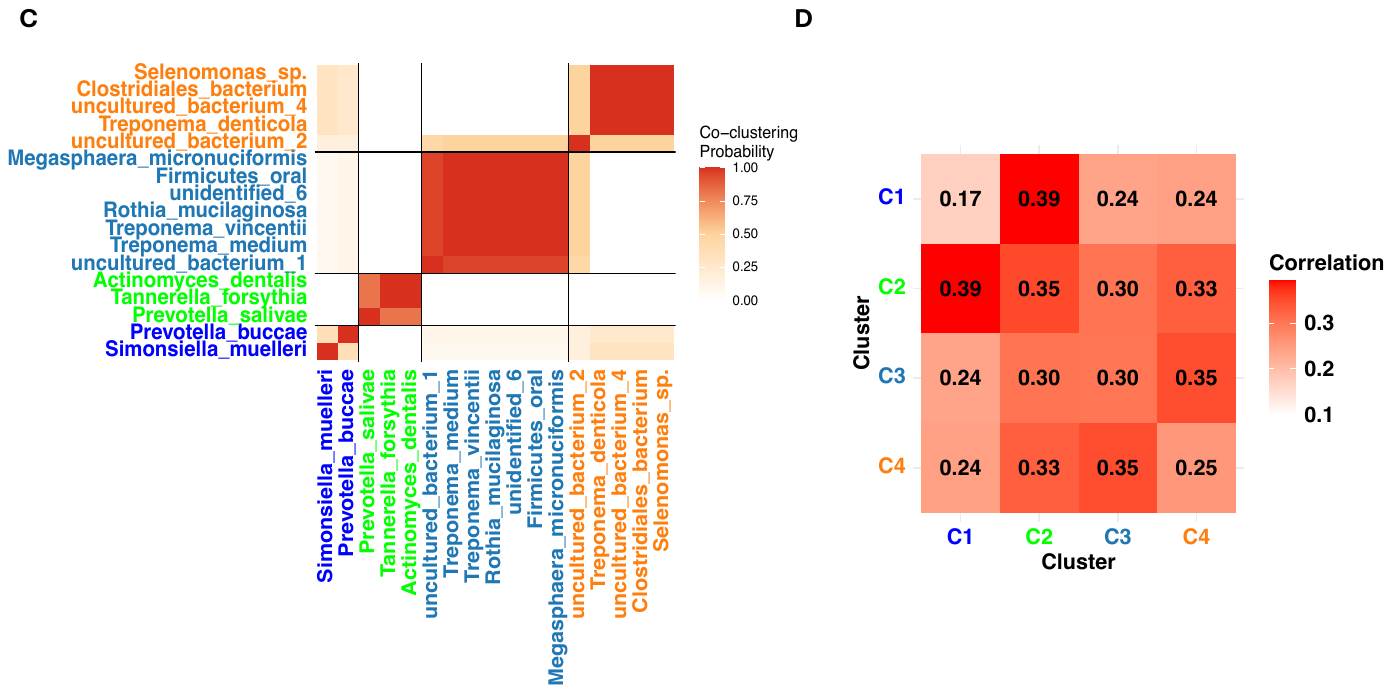}

    \caption{\textbf{Post-SALSO constrained resampling summary for BRACE-selected features.}
    (A)Comparison of competing methods in terms of sparsity and predictive performance. The left panel shows the mean number of selected variables, and the right panel shows the mean prediction error (PE), with horizontal error bars representing one standard deviation across 30 replicates.
    (B) Posterior point estimates and corresponding credible intervals for features selected by BRACE in the post-SALSO constrained resampling analysis. Point colours indicate SALSO-estimated cluster membership.
    (C) Posterior co-clustering probability matrix (posterior similarity matrix) for the selected features, ordered by the SALSO cluster labels.
    (D) Mean phylogenetic correlation matrix summarising within-cluster and between-cluster phylogenetic correlations among the selected features, aggregated according to the SALSO-derived clusters.
    }
\label{fig:Demmer_combined}
\end{figure}

We randomly divided the 111 samples into a training set of 83 samples and a test set of 28 samples, and fit the proposed and benchmarking models on the training data. The process was repeated for \textcolor{blue}{30} independent replicates, and we utilized these fitted models to calculate the prediction error on the test sets.  \textcolor{blue}{Summary metrics for prediction error and number of variables selected across 30 replicates are presented in Figure \ref{fig:Demmer_combined}, Panel A. Across all methods considered, BRACE consistently achieved the lowest prediction error and selected a stable number of features.
This sample-splitting approach supports the utility of our proposed method in accurately identifying true patterns of microbiome association, and is consistent with our simulation results in that the proposed method achieves the lowest prediction error across the methods considered.} 


To gain further insights into the role of oral microbiome composition in regulating insulin levels, we fit the proposed model on the full data. 
{\color{blue} Convergence of the MCMC sampler was assessed by running two independent chains for 100{,}000 iterations, discarding the first 50{,}000 as burn-in and thinning every 10th draw. Trace plots for the scalar parameters \((\sigma^2,\gamma^2,\alpha)\), the number of active clusters \((K_{\mathrm{active}})\), and the sparsity level \((1- \hat{\psi}_0)\) show good mixing and overlap between chains (Supplementary Figure S3), with Gelman--Rubin \(\hat{R}\) values between 1.00 and 1.02 for all monitored quantities, indicating satisfactory convergence.
The BRACE MCMC sampler jointly updates cluster assignments and cluster-level
parameters at each iteration, so the number and identity of clusters vary across the
posterior. To obtain a stable number of clusters, we apply the SALSO algorithm \cite{Dahl2022} to
the posterior draws of the partition, minimizing expected Variation of Information (VI)
loss. We select the number of active clusters $K$ via an elbow rule, increasing $K$
until the reduction in expected loss falls below $\delta = 0.01$.
}
Clustering stability and alignment were then evaluated by analyzing the posterior similarity (co-clustering) matrix (PSM), which quantifies how frequently pairs of observations are assigned to the same cluster over multiple iterations of our proposed algorithm. In Bayesian clustering methods, where label switching can cause variability in cluster assignments, the co-clustering matrix offers a robust measure of stability. By averaging these matrices, we derive a consensus partition that highlights stable groupings of label assignments. In Figure \ref{fig:Demmer_combined}, Panel A, we present a heatmap of the mean PSM after aligning the cluster labels using SALSO with four clusters. The dark diagonal blocks indicate that each cluster is consistently identified across iterations, reflecting high intra-cluster stability. Six clusters (including null and below threshold non-null) were selected with an expected VI loss of 0.42, as this partition best preserved intra-cluster cohesion while accommodating the inherent uncertainty in label assignments.


\textcolor{blue}{Next, to achieve FDR controlled variable selection, we used the approach described in Section \ref{ssec:VSFDR}.
BRACE identified 17 features grouped into four nonzero clusters, as illustrated in Figure \ref{fig:Demmer_combined} Panel B. It selected species belonging to the phyla Firmicutes and Bacteroidota to be associated with insulin levels. At the genus level, BRACE identified a number of species belonging to \textsl{Prevotella} and \textsl{Treponema}. As a validation check, we compared each selected species' Spearman screening correlation ($\rho$) with its posterior model coefficient ($\hat{\beta}$) (Supplementary Figure S4). Of the 17 selected species, 16 (94\%) showed concordant direction (binomial $p=0.0001$), confirming that the multivariable compositional model recovers genuine marginal signals rather than producing artifacts of the joint modeling. The single discordant species, \textit{Prevotella salivae}, had a near-zero screening correlation, suggesting its positive model coefficient reflects a conditional association that may have emerged only after adjusting for correlated taxa. Importantly, BRACE also identifies the shared-effect structure revealed by the taxa: species within the same cluster  occupy the same region of the concordance plot, confirming that the data-adaptive groupings capture meaningful similarities in how these taxa relate to insulin resistance.}

We now discuss the scientific findings in more detail, focusing on the bacterial species that confer increased risk. A cluster of three species was identified as having the strongest positive association with increased insulin levels. Within this cluster, \textsl{Tannerella forsythia} 
has long been recognized as one of the bacteria that contribute to periodontitis \citep{Socr1998}; more recently, increased abundance of \textsl{T.\ forsythia} in the oral microbiome has been associated with higher fasting blood glucose levels \citep{Chang2023}. 
This cluster also included a species belonging to the genus \textsl{Prevotella}, 
which supports the general understanding that 
\textsl{Prevotella} species contribute to increased inflammation \citep{kononen2022prevotella}.
Traditionally,  \textsl{Tannerella forsythia} along with \textsl{Treponema denticola} and \textsl{Porphyromonas gingivalis} species are believed to play a pathogenic role in periodontitis. However, in our analysis, \textsl{T. denticola} exhibited a negative association with insulin resistance, while \textsl{P. gingivalis} showed no association, mirroring observations in the ORIGINS study.
These findings could be data specific or suggest that the abundance of these species may rise as a response to host changes, such as overt hyperglycemia or periodontal disease, becoming relevant to systemic inflammation and insulin resistance only at more advanced stages of periodontal disease and dysglycemia \citep{demmer2017subgingival}.


\subsection{Cluster properties in terms of function and phylogeny}

We obtained a final clustering from the sampled cluster labels for our model using SALSO  as described in Section \ref{subsec::Prediction and selection results}. 
To assess whether the clusters obtained from BRACE represented functionally distinct characteristics, we conducted further analysis on the functional abundances of the taxa, obtained using PICRUSt2 \citep{douglas2020picrust2}, within each cluster. \textcolor{blue}{Principal Coordinates Analysis (PCoA) analysis of the functional feature distributions across the four clusters identified by BRACE revealed notable differences in functional abundances \cite{gower1966some}. This observation was further supported by Permutational Multivariate Analysis of Variance (PERMANOVA) performed on the distance matrices, which yielded a significant p-value $(p$ value $< 0.01)$, indicating statistically significant variation in functional abundances between clusters (Supplementary Figure S5) \cite{anderson2001new}.} A detailed explanation of the steps involved in functional analysis is provided in Supplementary Section S6.2.

Our next goal was to investigate whether these taxa clusters demonstrate phylogenetic similarity. To achieve that goal, we first constructed a phylogenetic tree based on the representative sequences for the observed taxa. Next, we calculated the phylogenetic correlation matrix $\mathbf{R}$ using the formula $ r_{ij} = \frac{l_{ij}}{\sqrt{l_{ii}}\sqrt{l_{jj}}}$, where $l_{aa}$ is defined as the branch length from the leaf node $a$ to the root node, for $a = 1, \dots, p$, and $l_{ij}$ is the shared branch length between leaf nodes $i$ and $j$. This matrix was calculated using the R package \texttt{ape} \citep{ape} and is illustrated in Figure S6. We then applied hierarchical clustering on the correlation matrix to understand the overlap between our estimated cluster labels and a clustering based on phylogenetic similarity. Figure S7 demonstrates that there is a small degree of overlap between the BRACE-determined cluster labels and the phylum-level groupings in the oral microbiome. This suggests that the groupings learned from BRACE may offer additional insight on features with similar functional effects, and do not simply recapitulate known taxonomy.

\textcolor{blue}{Next, given the estimated cluster labels, we calculated the within-cluster and between-cluster mean phylogenetic correlations, shown in Figure \ref{fig:Demmer_combined}: Panel D. We assessed phylogenetic coherence by comparing each cluster's within-cluster phylogenetic correlation to its three between-cluster correlations. Cluster C2 exhibits the strongest phylogenetic coherence, with its within-cluster correlation (0.35) exceeding two of three between-cluster values (0.30 with C3 and 0.33 with C4), indicating that this group captures taxa sharing both evolutionary history and similar effects on the outcome. Clusters C3 and C4 show mixed patterns: C3 has a within-cluster correlation of 0.30 that exceeds the between-cluster value with C1 (0.24) but not with C4 (0.35), while C4's within-cluster correlation (0.25) similarly exceeds only the C1 comparison (0.24). By contrast, cluster C1 is phylogenetically diverse, with a within-cluster correlation (0.17) lower than all between-cluster values, yet these taxa were grouped together by the model on the basis of their shared functional association with fasting insulin. 
This pattern directly reflects the motivating observation in Section \ref{sec:Intro} (Figure \ref{fig:motivation_fig}) that species-level associations with insulin are not conserved along phylogenetic lines, and closely related taxa can exhibit opposing effect directions. Rather than forcing all clusters to mirror the phylogeny, BRACE identifies phylogenetically coherent groups when such structure is present, while simultaneously discovering functionally convergent groupings among phylogenetically distant taxa. Taken together, these results demonstrate that the clusters identified by our method reflect shared outcome-relevant effects that go beyond what phylogenetic proximity alone would predict, underscoring the value of data-adaptive aggregation over tree-based approaches.}

\subsection{\textcolor{blue}{Gut microbiome in obesity}} \textcolor{blue}{To showcase the broad applicability of our method, we include an additional case study on the association of the gut microbiota with body mass index (BMI) in Supplementary Section S7.}

\section{Conclusion} \label{sec:conclusion}
\textcolor{blue}{In this article, we present a Bayesian nonparametric approach for microbiome compositional regression that performs dimension reduction through data-adaptive clustering of shared regression effects. The proposed methodology advances regression modeling for microbiome data in two key respects. First, it introduces a Bayesian nonparametric prior construction for high-dimensional
compositional regression that enables flexible microbiome effect aggregation and simultaneous feature selection}.
\textcolor{blue}{Second, we develop a projection-based prior construction that enforces the compositional log-contrast constraint directly on the regression coefficients. Through simulation studies and real-data applications, we show that the proposed model can improve estimation, prediction, and feature-selection performance relative to existing compositional regression methods.}

Our proposed method can be extended to incorporate additional covariates, which could be relevant for many health-related studies. To accommodate additional clinical or demographic variables we can write $ \boldsymbol{y} = \mathbf{X}\boldsymbol{\beta} + \mathbf{C}\boldsymbol{\zeta} + \boldsymbol{\varepsilon}$, where $\mathbf{C}$ is the matrix of additional covariates and  $\boldsymbol{\zeta} \sim N(\boldsymbol{0}, \sigma_{\zeta}^2\mathbf{I})$. In addition, our formulation can be extended to model binary or count responses by changing the linear link function to probit/logit or log, but these extensions are non-trival and will be explored in future work. 

We acknowledge that microbiome datasets often contain a high proportion of zeros. \textcolor{blue}{Although pseudocount-based preprocessing may introduce bias in log-contrast models, our results suggest that the conclusions are robust across the zero-handling strategies considered here.} We do not explicitly address the handling of zeros within our model structure, which is typical for methods considering microbiome features as predictor variables \citep{lin2014variable, Shi2016, zhang2024bayesian}. Explicit modeling of zeros is more common in differential abundance analysis or when the microbiome is treated as a response, where zero-inflation can be integrated into the assumed data distribution \citep{zhang2015zero, lee2020bayesian, koslovsky2023bayesian}.


\textcolor{blue}{Finally, a main limitation of the proposed approach is its computational cost, particularly in the cluster-label updates, which require repeated evaluation of collapsed marginal likelihoods involving matrix inversions over the active clusters. As a result, the collapsed Gibbs sampler is most efficient when the number of inferred clusters \(K\) is small relative to the number of predictors \(p\). In settings where the posterior favors a large number of clusters, the sampler may become computationally burdensome. For this reason, we recommend initializing the sampler with a modest number of clusters and monitoring the posterior behavior of \(K\). Future work could improve scalability by developing approximate inference strategies, such as variational Bayes or other deterministic approximations, for larger microbiome data sets with many active clusters.}

\section*{Acknowledgments}
 We would like to express our gratitude to Dr.\ Yushu Shi for her invaluable feedback on an earlier draft of this paper. S.S.\ was partially supported by NIH R01 HL158796. K.A.D.\ was partially supported by NIH/NCI CCSG P30CA016672, CCTS TR000371 and CPRIT RP160693. C.B.P.\ was partially supported by NIH R01 HL158796, NIH/NCI CCSG P30CA016672, and an Andrew Sabin Family Fellowship. The author(s) acknowledge the support of the High Performance Computing for research facility at the University of Texas MD Anderson Cancer Center for providing computational resources that have contributed to the research results reported in this paper.


\section{Supplementary Material}
\begin{description}

\item[Supplementary Results ] Appendices, Tables, and Figures are available in the Supplementary Materials. (pdf file)

\item[R Code for BRACE ] The R code for an illustrative example presented in this article, along with the code for reproducing the figures included in the main manuscript is available as a zipped tar file as part of the Supplementary Material. The R code is also publicly accessible at \href{https://github.com/satabdisaha1288/BRACE}{https://github.com/satabdisaha1288/BRACE}.

\item[Real Dataset ] Our case study data (ORIGINS data set) was provided in association with \cite{demmer2017subgingival} and \cite{marotz2022early}. The fastQ files can be accessed from  the European Nucleotide Archive (ENA, \href{https://www.ebi.ac.uk/ena/browser/home}{https://www.ebi.ac.uk/ena/browser/home}) using Project reference ID PRJEB50306. The metadata and processed sequences can be downloaded from \href{https://qiita.ucsd.edu}{Qiita.ucsd.edu} using the study ID
11808. The fastQ files from Qiita were processed using Qiime 2 for obtaining the taxa abundances \citep{bolyen2019reproducible}.

\end{description}


\bibliographystyle{plain}
\bibliography{library}

\end{document}